\documentclass[AEJ]{AEA}
\usepackage{amssymb}
\usepackage{amsfonts}
\usepackage{amsmath}
\usepackage{amstext}
\usepackage{graphicx}
\usepackage{natbib}
\usepackage{setspace}
\draftSpacing{1.5}
\usepackage{hyperref}
\usepackage{xcolor}
\hypersetup{
	colorlinks,
	linkcolor={red!50!black},
	citecolor={blue!50!black},
	urlcolor={blue!80!black}
}

\newtheorem{lemma}{\it\textsc{Lemma}}[section]

\newtheorem{theorem}{\it\textsc{Theorem}}[section]
\newtheorem{corollary}{\it\textsc{Corollary}}[section]
\newtheorem{remark}{Remark}[section]

\newtheorem{definition}{Definition}[section]

\begin{document}

\title{Sovereign Default Risk and Uncertainty Premia}
\shortTitle{Default and Uncertainty Premia}
\author{Demian Pouzo and Ignacio Presno \thanks{Pouzo: Department of Economics, UC at Berkeley, 530 Evans Hall \# 3880,
Berkeley, CA 94720-3880. E-mail: dpouzo@econ.berkeley.edu. Presno: Department of Economics, Universidad de Montevideo, 2544 Prudencio de Pena St., Montevideo, Uruguay 11600. Email: jipresno@um.edu.uy.
	We are deeply grateful to Thomas J. Sargent for his constant guidance and
	encouragement. We also thank Fernando Alvarez, David Backus, Timothy Cogley, Ricardo Colacito, Ernesto dal Bo, Bora Durdu, Ignacio Esponda, Gita Gopinath, Yuriy Gorodnichenko, Juan Carlos Hatchondo, Lars Ljungqvist, Jianjun Miao, Anna Orlik, Viktor Tsyrennikov and Stanley Zin for helpful comments.}}
\date{First Version: December 18, 2010. This Version: \today}

\pubMonth{Month}
\pubYear{Year}
\pubVolume{Vol}
\pubIssue{Issue}
\JEL{D81, E21, E32, E43, F34.}
\Keywords{sovereign debt, default risk, model uncertainty, robust control.}

\begin{abstract}
This paper studies how international investors' concerns about model misspecification affect sovereign bond spreads. We develop a general equilibrium model of sovereign debt with endogenous default wherein investors fear that the probability model of the underlying state of the borrowing economy is misspecified. Consequently, investors  demand higher returns on their bond holdings to compensate for the default risk in the context of uncertainty. In contrast with the existing literature on sovereign default, we match the bond spreads dynamics observed in the data together with other business cycle features for Argentina, while preserving the default frequency at historical low levels.
\end{abstract}

\maketitle

\section{Introduction}

Sovereign defaults, or debt crises in general, are a pervasive economic phenomenon, especially among emerging economies. Recent defaults by Russia (1998), Ecuador (1999) and Argentina (2001), and the ongoing debt crisis of Greece and other peripheral eurozone countries have put sovereign default issues at the forefront of economic policy discussion. A central issue for the assessment of any policy proposal hinges on how the borrowing costs and market conditions are expected to evolve once the policy is adopted. Therefore, constructing economic models that can provide accurate predictions in terms of pricing while generating seldom default events, is key.

As documented by several studies, however, a well-known puzzle in the sovereign default literature built on the general equilibrium framework of \cite{Eaton} is: why are these models unable to account for the observed dynamics of the bond spreads, while preserving the default frequency at historical low levels\footnote{This phenomenon is not limited to the sovereign debt literature, since it is also well-documented in the corporate debt literature; see \cite{Huang_Huang} and \cite{Elton}, for example.}? A possible explanation of the lack of success of these models when confronted with asset prices data may be attributed to the fact that they follow the rational expectation hypothesis: agents know the data-generating process, which coincides with their own subjective beliefs.

This paper tackles this ``pricing puzzle'' ---while also accounting for other salient empirical features of the real business cycles\footnote{For a summary of the empirical regularities in emerging economies, see, e.g., \cite{Neumeyer_Perri}.}--- by introducing international lenders that distrust their probability model governing the evolution of the state of the borrowing economy and want to guard themselves against specification errors in it. In doing so, we relax the rational expectations assumption by allowing the lenders to exhibit ``uncertainty aversion'', also commonly known as ``Knightian uncertainty''. In our model, a borrower (e.g., an emerging economy) can trade long-term bonds with international lenders in financial markets, similar to \cite{Chatterjee_2010}. Debt repayments cannot be enforced and the emerging economy may decide to default at any point of time. Lenders in equilibrium anticipate the default
strategies of the emerging economies and demand higher returns on their sovereign bond holdings to compensate for the default risk. In case of
default, the economy is temporarily excluded from financial markets and suffers a direct output cost. In this setting, we show how lenders'
desire to make decisions that are robust to misspecification of the conditional probability of the borrower's endowment alters the returns on sovereign bond holdings.\footnote{By following \cite{Eaton}, we abstract from transaction costs, liquidity restrictions, and other frictions that may affect the real return on sovereign bond holdings.}

The assumption about concerns about  model misspecification is intended to
capture the fact that international lenders may distrust their statistical model used to predict relevant macroeconomic variables of the emerging economy. Alternatively, lenders could be aware of the limited availability of
\emph{reliable} official data.\footnote{\cite{Boz} document the availability of significantly shorter time series for most relevant economic indicators in emerging economies than in developed ones. For example, in the database from the International Financial Statistics of the IMF the median length of available GDP time series at a quarterly frequency is 96 quarters in emerging economies, while in developed economies is 164.} This issue has become more severe in recent years in some emerging economies, particularly in Argentina, where the government's intervention in the computation of the consumer price index is known worldwide, motivating warning calls for correction coming from international credit institutions. By under-reporting inflation, the Argentinean government has been over-reporting real GDP growth. Concerns about model misspecification can also be attributed to measurement errors, and lags in the release of the
official statistics together with subsequent revisions. These arguments are aligned with the suggested view of putting the econometrician and the economic agent in a position with identical information, and limitations on their ability to estimate statistical models.

The novelty in our paper comes from the fact that lenders are uncertainty averse in the sense that they are unwilling or unable to fully trust a unique probability distribution or \emph{probability model} for the endowment of the borrower, and at the same time dislike making decisions in the context of multiple probability models. To express these doubts about model uncertainty, following \cite{HS_WP05} we endow lenders with
multiplier preferences.\footnote{Axiomatic foundations for this class of preferences have been provided by \cite{Strzalecki}.}$^{,}$
Lenders in our model share a \emph{reference} or \emph{approximating} probability model for the borrower's endowment, which is their best estimate of the economy's dynamics. They acknowledge, however, that it may be misspecified, and express their doubts about model misspecification by contemplating alternative probability distributions that are statistical perturbations of the reference probability model. To make choices that perform well over this set of probability distributions, the lender \emph{acts as if} contemplating a conditional  worst-case probability that is distorted relative to his approximating one. This distorted distribution therefore arises from perturbing the approximating model by slanting probability towards the states associated with low utility. In our model, these low-utility states for the lender coincide with those in which the payoff of the sovereign bond is lower, because default occurs in the first place or the market value of the outstanding debt drops.

The main result of our paper is that by introducing lenders' fears about model
misspecification our calibration matches the high, volatile, and typically countercyclical bond
spreads observed in the data for the Argentinean economy, together with
standard business cycle features while keeping the default frequency
at historical levels. At the same time, our model can account for the average risk-free rate observed in the data. Interestingly, we find that if the borrower can issue long-term debt model uncertainty almost does not affect quantitatively its level of indebtness. It is worth pointing out that in the simulations we also find that in a simple rational expectations framework, for plausible values of the parameters, risk aversion alone on the lenders' side with time-separable CRRA preferences is not sufficient to generate the observed
risk premia; to some extent, this is an analogous result to the equity premium puzzle studied in \cite{Mehra_Prescott}. Additionally, as the degree of lenders' risk aversion increases, the average net risk-free rate declines, eventually to negative levels.

The intuition behind our results is as follows. Under the assumption
that international lenders are risk neutral and have rational expectations (by fully trusting the data generating process), as for example in \cite{ARELLANO_AER08}, the equilibrium prices of long-term bonds are simply given by the present value of adjusted conditional probabilities of
not defaulting in future periods. Consequently, the
pricing rule in these environments prescribes a strong connection between
equilibrium prices and default probability. When calibrated to the data,
matching the default frequency to historical levels (the consensus number
for Argentina is around $3$ percent annually), delivers spreads that are too
low relative to those observed in the data.\footnote{\cite{ARELLANO_AER08}, \cite{Lizarazo}, and \cite{Hatchondo_Debt_Dilution}, use a default frequency of 3 percent per year. \cite{Yue} and \cite{Mendoza_Yue} target an annual default frequency of 2.78 percent. Also, \cite{Reinhart} finds that emerging economies with at least one episode of external default or debt restructuring defaulted roughly speaking three times every $100$ years over the period from $1824$ to $1999$.} Our methodology breaks this strong connection by introducing a different probability measure, the one in which lenders' uncertainty aversion is manifested. In our case, there is a strong connection between equilibrium prices and the default probability under this new worst-case probability measure. 

From an asset pricing perspective, the key element in generating high spreads while
matching the default frequency is a sufficiently negative correlation of the market stochastic
discount factor with the payoff of the bond. With fears about model misspecification, the stochastic discount factor has an additional component given by the probability distortion inherited in the worst-case density for the borrower's endowment. This probability distortion, which is low when the borrower repays and particularly high when the borrower defaults or the market value of the outstanding debt falls, induces in general the desired negative co-movement between the stochastic discount factor and the payoff of the bond. Some recent papers, such as \cite{ARELLANO_AER08}, \cite{Arellano_WP}, and \cite{Hatchondo_Debt_Dilution}, assume instead an ad hoc functional form for the market stochastic discount factor in order to generate sizable bonds spreads as observed in the data. Our paper can therefore be seen as providing microfoundations for valid stochastic discount factors.


In our model with a defaultable asset, this endogenous probability distortion is non-smooth in the realization of the borrower's next-period endowment and exhibits an inverse V-shaped kink due to the default contingency. This yields an endogenous hump of the worst-case density over the interval of endowment realization in which default is optimal. This special feature is unique to this current setting. A direct implication of this is that the subjective probability assigns a significantly higher probability to the default event than the actual one. \footnote{ We can view the default event as a ``disaster event'' from the lenders' perspective, this result links to the growing literature on ``rare events''; see, for example, \cite{BARRO_QJE06}.} Fears about model misspecification then amplify its effect on both allocations and equilibrium prices, as they increase the lenders' perceived likelihood of these events occurring. As the latter are typically higher in the model than the default probability implied by the best proxy of the economic dynamics given the data limitations, our story is therefore in line with the commonly known ``peso problems''. We find this an interesting contribution of our paper.

We also show analytically in our benchmark model that the relative size of the endowment for the lender does not affect the equilibrium bond prices or the borrower's allocations. Besides the theoretical contribution, this result implies that there is no need to identify who the lenders are in the data, and, in particular, to find a good proxy of their income relatively to the borrower's endowment.

\textbf{Related Literature.} This paper builds on and contributes to two
main strands of the literature: sovereign default, and robust control theory and ambiguity aversion or Knightian uncertainty, in
particular applied to asset pricing.

\cite{ARELLANO_AER08} and \cite{AG_1} were the first to extend \cite{Eaton}
general equilibrium framework with endogenous default and risk neutral
lenders to study the business cycles of emerging economies. \cite{Chatterjee_2010} introduced long-term debt in these environments.\footnote{From a technical perspective, \cite{Chatterjee_2010} proposes an alternative
approach to handle convergence issues. The authors consider an i.i.d. output
shock drawn from a continuous distribution with a very small variance. Once
this i.i.d. shock is incorporated, they are able to show the existence of a
unique equilibrium price function for long-term debt with the property that
the return on debt is increasing in the amount borrowed.} \cite{Lizarazo} endows the lenders with constant relative risk aversion (CRRA)\ preferences.
\cite{Verdelhan} has studied the setup with positive correlation between
lenders' consumption and output in the emerging economy in addition to
time-varying risk aversion on the lenders' side as a result of habit formation.

To our knowledge, the paper that is the closest to ours is the independent work by \cite{Costa}.
That paper also assumes that international lenders want to guard themselves
against specification errors in the stochastic process for the endowment of the borrower, but this is
achieved in a different form. In our model, lenders are endowed with \cite%
{HS_WP05} multiplier preferences.
In contrast, in \cite{Costa}%
\ the worst-case density minimizes the expected value of the bond. Moreover,
\cite{Costa} considers one-period bonds and assumes lenders live for one period only.

Other recent studies that have focused on business cycles in emerging economies in the presence of fears about model misspecification are \cite{Young} and \cite{LNY}. \cite{Young} studies optimal tax policies to deal with sudden stops when households and/or agents distrust the stochastic process for tradable total factor productivity shocks, trend productivity, and the interest rate. \cite{LNY} explores the role of robustness and information-processing constraints (rational inattention) in the joint dynamics of consumption, current account, and output in small open economies.

Finally, our paper relates to the growing literature analyzing the asset-pricing implications of ambiguity. \cite{BHS} finds that introducing concerns about robustness to model misspecification can yield combinations of the market price of risk and the risk-free rate that approach \cite{Hansen_Jagannathan} bounds. Using a dynamic portfolio choice problem of a robust investor, \cite{Maenhout} can explain high levels of the equity premium, as observed in the data. \cite{Fragile_Beliefs} generates time-varying risk premia in the context of model uncertainty with hidden Markov states. \footnote{In a consumption-based asset-pricing model, \cite{Boyarchenko} studies the dynamics of the CDS spreads by contemplating uncertainty about the signal-extraction process and the underlying economic model. \cite{Miao} considers a pure-exchange economy with hidden Markov regime-switching processes for consumption and dividends, with agents with generalized recursive smooth preferences, closely related to \cite{Klibanoff} model of preferences. \cite{ES} studies the impact of uncertain information quality on asset prices in a model of learning with investors endowed with recursive multiple-priors utility, axiomatized in \cite{ES_2003}. More asset-pricing applications with different formulations of ambiguity aversion are \cite{Epstein_Wang}, \cite{Chen_Epstein}, \cite{Hansen}, \cite{Bidder_Smith} and \cite{Drechsler}.}

\textbf{Roadmap.} The paper is organized as follows. Section \ref{sec:model} presents the model. In Section \ref{sec:EQ_T1} we introduce the recursive equilibrium in our economy and describe the implications of model uncertainty on equilibrium prices. In Section \ref{sec:quant} we calibrate our model to Argentinean data and present our quantitative results for long-term bonds. Section \ref{sec:dep} disciplines the degree of robustness in our economy using detection error probabilities and a new moment-based uncertainty measure. 
 Finally, Section \ref{sec:conc} concludes.

\section{The Model}
\label{sec:model}

In our model an emerging economy interacts with a continuum of identical international
lenders of measure 1. The emerging economy is populated by a representative,
risk-averse household and a government.

The government in the emerging economy can trade a long-term bond with atomistic international lenders to smooth
consumption and allocate it optimally over time. Throughout the paper we will refer to the emerging
economy as the \textit{borrower}. Debt contracts cannot be enforced and the
borrower may decide to default at any point of time. In case the government defaults on its debt, it incurs two types of costs. First, it is temporarily excluded from financial markets. Second, it suffers a direct
output loss.

While the borrower fully trusts the probability model governing the evolution
of its endowment, which we will refer to as the
\textit{approximating model}, the lender suspects it is misspecified. From here on, we will use the terms \emph{probability model} and \emph{distribution} interchangeably. For this reason, the lender contemplates a set of alternative
models that are statistical perturbations of the approximating model, and
wishes to design a decision rule that performs well across this set of
distributions.\footnote{In order to depart as little as possible from \cite{Eaton}
framework, throughout the paper we assume that the lender distrusts only
the probability model dictating the evolution of the endowment of the
borrower, not the distribution of any other source of uncertainty, such as the random variable that indicates whether the borrower re-enters financial markets or not. At the same time, and for the same reason, we assume the extreme case of no doubts about model misspecification on the borrower's side.}

\subsection{Stochastic Process of the Endowment}

Time is discrete $t=0,1,\ldots $. Let $(W_{t})_{t=0}^{\infty} \equiv (X_{t},Y_{t})_{t=0}^{\infty}$ be an stochastic process describing the borrower's endowment. In particular, let $(Y_{t})_{t=0}^{\infty}$ be a discrete-state Markov Chain, $(\mathbb{Y},P_{Y^{\prime}|Y},\nu)$ where $\mathbb{Y} \equiv \{y_{1},...,y_{|\mathbb{Y}|}\} \subseteq \mathbb{R}_{+}$, $P_{Y^{\prime}|Y}$ is the transition matrix and $\nu$ is the initial probability measure, which is assumed to be the (unique) invariant (and ergodic) distribution of $P_{Y^{\prime}|Y}$. Let $(X_{t})_{t=0}^{\infty}$ be such that, for all $t$, $X_{t} \in [\underline{x},\bar{x}] \equiv \mathbb{X} \subseteq \mathbb{R}$ is an i.i.d. continuous random variable, i.e., $X_{t} \sim P_{X}$ and $P_{X}$ admits a pdf (with respect to Lebesgue), which we denote as $f_{X}$. Henceforth, we define $\mathbb{W} \equiv \mathbb{X} \times \mathbb{Y}$ and $P_{W'|W}$ denotes the conditional probability of $W_{t+1}$, given $W_{t}$, given by the product of $P_{Y^{\prime}|Y}$ and $P_{X}$; $P$ denotes the probability, induced by $P_{W'|W}$ over infinite histories, $w^{\infty} = (w_{0},...,w_{t},...)$; finally, we also use $\mathcal{W}^{t}$ to denote the
$\sigma$-algebra generated by the partial history $W^{t} \equiv
(W_{0},W_{1},...,W_{t})$.

The reason behind our definition of $W_{t}$ will become apparent below, but, essentially, we think of $Y_{t}+X_{t}$ as the borrower's endowment at time $t$, and the separation between $Y_{t}$ and $X_{t}$ is due to numerical issues that appear in the method for solving the model; see \cite{Chatterjee_2010} for a more thorough discussion.

Finally, we use $\overline{z}$ to denote the endowment of the lender,
which is chosen to be non-random and constant over time for
simplicity.




\begin{remark}
	Throughout the paper, for a generic random variable $W$, we use $W$ to denote the random variable and $w$ to denote a particular
	realization. Except for bond holdings, where we use $B$ for the government's and $b$ for the lenders bond holdings.
\end{remark}

\subsection{Timing Protocol}

We assume that all economic agents, lenders, and
the government (which cares about the consumption of the representative
household), act sequentially, choosing their allocations period by period.

The economy can be in one of two stages at the beginning of each period $%
t $: financial autarky or with access to financial markets.

The timing protocol within each period is as follows. First, the endowments
are realized. If the government has access to financial markets, it decides whether to
repay its outstanding debt obligations or not. If it decides to repay, it chooses new bond holdings
and how much to consume. Then, atomistic international lenders---taking prices as
given---choose how much to save and how much to consume. The minimizing agent, who
is a metaphor for the lenders' fears about model misspecification, chooses
the probability distortions to minimize the lenders' expected utility. Due
to the zero-sumness of the game between the lender and its minimizing agent,
different timing protocols of their moves yield the same solution. If the
government decides to default, it switches to
autarky for a random number of periods.
While the government is excluded from financial markets, it has no decision to make and simply awaits re-entry to financial markets.

\subsection{Sovereign Debt Markets}

Financial markets are incomplete. Only a non-contingent, long-term bond can be traded between the borrower and the lenders. The borrower, however, can default on this bond at any time, thereby adding some degree of state contingency.

As in \cite{Chatterjee_2010}, the long-term bond exhibits a simplified payoff structure. We assume that in each period a fraction $\lambda$ of the bond matures, while a coupon $\psi$ is paid off for the remaining fraction $1-\lambda$, which is carried over into next period; $\psi$ and $\lambda$ are primitives in our model. Modeling the bond this way is convenient to keep the problem tractable by avoiding too many state variables. Under these assumptions, it is sufficient to keep track of the outstanding quantity of bonds of the borrower to describe his financial position.

Bond holdings of the government and of the individual lenders, denoted by $B_{t}\in \mathbb{B} \subseteq \mathbb{R}$ and $b_{t}\in \mathbb{B}\subseteq \mathbb{R}$, respectively, are $\mathcal{W}^{t-1}$-measurable. The set $\mathbb{B}$ is bounded and thereby includes possible borrowing or savings limits.\footnote{Positive bond holdings $B_{t}$ means that the government enters period $t$ with net savings, that is, in net term it has been purchasing bonds in the past. }


The borrower can choose a new quantity of bonds $B_{t+1}$\ at a price $q_{t}$. A
debt contract is given by a vector $(B_{t+1},q_{t})$ of quantities of bonds
and corresponding bond prices. The price $q_{t}$\ depends on the borrower's
demand for debt at time $t,$\ $B_{t+1},$\ and his endowment $y_{t}$, since these variables affect his incentives to default. In this class of models, generally, the higher the level of indebtness and/or the lower the (persistent) borrower's endowment, the
greater the chances the borrower will default (in future periods) and, hence,
the lower the bond prices in the current period.

For each $y \in \mathbb{Y}$, we refer to $q(y,\cdot ):\mathbb{B} \rightarrow \mathbb{R}_{+}$
as the \emph{bond price function}.\footnote{As we show below, the bond price function only depends on $(y_{t},B_{t+1})$, and not on $x_{t}$.} Thus, we can define the set of debt contracts
available to the borrower for a given $w$ as the graph of
$q(y,\cdot)$.\footnote{The graph of a function, $f: \mathbb{X} \rightarrow \mathbb{Y}$, is the set of $\{(x,y) \in \mathbb{X} \times \mathbb{Y}
  \colon y = f(x)~and~x\in \mathbb{X} \}$.}

\subsection{Borrower's Problem}
\label{sec:BORROWERS_BENCHMARK}

The representative household in the emerging economy derives utility from
consumption of a single good in the economy. Its preferences over
consumption plans can be described by the expected lifetime utility\footnote{A consumption plan is a stochastic process, $(c_{t})_{t}$, such that $c_{t}$ is $\mathcal{W}^{t}$-measurable.}
\begin{equation}
E \left[\sum_{t=0}^{\infty }\beta ^{t}u(c_{t}) |w_{0} \right], \label{HH util}
\end{equation}
where $E \left[ \cdot |w_{0} \right]$ denotes the expectation under the probability measure $P$ (conditional on
time zero information $w_{0}$), $\beta \in (0,1)$ denotes the time discount
factor, and the period utility function $u:\mathbb{R}_{+} \rightarrow \mathbb{R}$ is strictly increasing and strictly concave, and satisfies the Inada conditions.\footnote{Note that the assumption that the representative household and the government fully trust the approximating model $P$ is embedded in $E \left[ \cdot |w_{0} \right]$.}

The government in this economy, which is benevolent and maximizes the
household's utility (\ref{HH util}), may have access to international
financial markets, where it can trade a long-term bond with the
foreign lenders. While the government has access to the financial markets, it can sell or
purchase bonds from the lenders and make a lump-sum transfer across
households to help them smooth consumption over time.



For each $(w_{t},B_{t})$, let $V(w_{t},B_{t})$ be the value (in terms of lifetime
utility) for the borrower of having the option to default, given an
endowment vector $w_{t},$ and outstanding bond holdings equal to $B_{t}$. Formally, the borrower's value of having access to financial markets $V(w_{t},B_{t})$ is given by
\begin{equation*}
V(w_{t},B_{t})=\max \left\{
V_{A}(\underline{x},y_{t}),V_{R}(w_{t},B_{t})\right\},
\end{equation*}%
where $V_{A}(x_{t},y_{t})$ is the value of exercising the option to default, given an endowment vector $w_{t}=(y_{t},x_{t})$, and $V_{R}(w_{t},B_{t})$ is the value of repaying the outstanding debt, given state $(w_{t},B_{t})$. In the period announcing default, the continuous component of endowment $x_{t}$ drops to its lowest level $\underline{x}$. For the rest of the autarky periods, however, $x_{t}$ is stochastic and drawn from the distribution $P_{X}$, mentioned before. Throughout the paper we use subscripts $A$ and $R$ to denote the values
for \textit{autarky} and \textit{repayment}, respectively.

Every period the government enters with access to financial markets, it evaluates the present lifetime
utility of households if debt contracts are honored against the present
lifetime utility of households if they are repudiated. If the former
outweighs the latter, the government decides to comply with the contracts,
makes the principal and coupon payments for the debt carried from the last period $B_{t}$, totaling $(\lambda + (1-\lambda)\psi)B_{t}$, and chooses next period's bond holdings $B_{t+1}$. Otherwise, if the utility of defaulting on
the outstanding debt and switching to financial autarky is higher, the government
decides to default on the sovereign debt.

Consequently, the value of repayment $V_{R}(w_{t},B_{t})$ is
\begin{align*}
V_{R}(w_{t},B_{t}) =&\max_{B_{t+1} \in \mathbb{B}}
u(c_{t})+\beta E \left[ V(W_{t+1},B_{t+1}) \mid w_{t} \right] \\
s.t.&~c_{t} = y_{t}+x_{t}-q(y_{t},B_{t+1})(B_{t+1}-(1-\lambda) B_{t}) +(\lambda + (1-\lambda)\psi)B_{t}.
\end{align*}

Finally, the value of autarky $V_{A}(w_{t})$ is
\begin{equation*}
V_{A}(w_{t})=u(y_{t} + x_{t} - \phi(y_{t}))+\beta E \left[ (1-\pi )V_{A}(W_{t+1})+\pi V(W_{t+1},0) \mid w_{t} \right],
\end{equation*}
where $\pi $ is the probability of re-entering financial markets next
period.\footnote{As in \cite{ARELLANO_AER08}, we do not model the exclusion from financial markets as an endogenous decision by the lenders. By modeling this punishment explicitly in long-term financial relationships, \cite{Kletzer_Wright} show how international borrowing can be sustained in equilibrium through this single credible threat.} In that event, the borrower enters next period carrying no debt, $B_{t+1}=0$.\footnote{%
Notice that we assume there is no debt renegotiation nor any form of debt restructuring mechanism.
\cite{Yue} models a debt renegotiation process as a Nash bargaining game played
by the borrower and lenders. For more examples of debt renegotiation, see \cite{Benjamin_Wright} and \cite{Pitchford_Wright}.
 \cite{Pouzo} assumes a debt restructuring mechanism
in which the borrower receives random exogenous offers to repay a fraction of the defaulted debt. A positive rate
of debt recovery gives rise to positive prices for defaulted debt that can be traded amongst lenders
in secondary markets.}
The function $\phi : \mathbb{Y} \rightarrow \mathbb{Y}$ such that $y \geq
\phi(y) \quad \forall y \in \mathbb{Y}$ represents an
ad hoc direct output cost on $y_{t}$, in terms of consumption units, that
the borrower suffers when excluded from financial markets. This output
loss function is consistent with evidence that shows that countries
experience a drop in output in times of default due to the lack of short-term
trade credit and financial disruption in the banking sector, among others.\footnote{\cite{Mendoza_Yue} endogenize this output loss as an outcome
that results from the substitution of imported inputs by less-efficient
domestic ones when the financing cost of the former increases along with the sovereign default risk.}
Notice that in autarky the borrower has no decision to make and simply
consumes $y_{t}  - \phi(y_{t}) + x_{t}$.

The default decisions are expressed by the indicator $\delta : \mathbb{W} \times \mathbb{B}%
\rightarrow \{0,1\}$, that takes value $0$ if default is optimal; and $1$, otherwise; i.e., for all $(x,y,B)$, $\delta (x,y,B)=\mathbf{1}\left\{ V_{R}(x,y,B)\geq V_{A}(\underline{x},y)\right\}$.

\subsection{Lenders' Preferences and their Fears about Model Misspecification}
\label{sec:LENDERS_PREF}

We assume that the lenders' have per-period payoff linear in consumption, while also being uncertainty averse or ambiguity averse. Since the i.i.d. component $x_{t}$ is introduced merely for computational purposes ---to guarantee convergence, as in \cite{Chatterjee_2010}---, we assume no doubts about the specification of its distribution.

The lenders distrust, however, the
probability model which dictates the evolution of $y_{t}$, given by the \emph{approximating model} $P_{Y^{\prime}|Y}$. For this reason, they contemplate a set
of alternative densities that are statistical perturbations of the
approximating model, and wish to design a decision rule that performs well
over this set of priors. These alternative conditional probabilities,
denoted by $\widetilde{P}_{Y,t}(\cdot|w^t)$ for all $(t,w^t)$, are assumed
to be absolutely continuous with respect to $P_{Y^{\prime}|Y}(\cdot | y_{t})$, i.e. for all $A \subseteq
\mathbb{Y}$ and $w^{t} \in \mathbb{W}^{t}$, if $P_{Y^{\prime}|Y}(A|y_{t}) = 0$, then $\widetilde{P}_{Y,t}(A|w^{t}) = 0$.\footnote{Note that the distorted probabilities $\widetilde{P}_{Y,t}$ do not necessarily inherit the properties of $P_{Y^{\prime}|Y}$, such as its Markov structure. At the same time, they may depend on the history of past realizations of all shocks, including $x^{t}$, as these may affect equilibrium allocations.}

In order to construct any of these distorted probabilities $\widetilde{P}_{Y,t}$, for each $t$, let $m_{t+1} : \mathbb{Y} \times \mathbb{W}^{t} \rightarrow \mathbb{R}_{+}$ be the \emph{conditional likelihood ratio}, i.e., for any $y_{t+1}$ and $w^{t}$,
\begin{equation*}
m_{t+1}(y_{t+1}|w^{t})=\left\{
\begin{array}{cc}
\frac{\widetilde{P}_{Y,t}(y_{t+1}|w^{t})}{P_{Y^{\prime}|Y}(y_{t+1}|y_{t})}
& \text{if }P_{Y^{\prime}|Y}(y_{t+1}|y_{t})>0 \\
1 & \text{if }P_{Y^{\prime}|Y}(y_{t+1}|y_{t})=0%
\end{array}%
\right. .
\end{equation*}
Observe that for any $(t,w^t)$, $m_{t+1}(\cdot|w^t) \in \mathcal{M}$ where:
 \begin{equation*}
 \mathcal{M} \equiv \{ g \colon\mathbb{Y} \rightarrow \mathbb{R}_{+} ~|~ \sum_{y' \in \mathbb{Y}}  g(y') P_{Y^{\prime}|Y}(y'|y) = 1,~\forall y \in \mathbb{Y} \}.
 \end{equation*}

Following \cite{HS_WP05} and references therein, to express fears about
model misspecification we endow lenders with multiplier preferences.
 While the
lenders choose bond holdings to maximize their utility, the minimizing agent
chooses a sequence of distorted conditional probabilities $(\widetilde{P}_{Y,t+1})_{t}$, or equivalently a sequence of conditional likelihood ratios $(m_{t+1})_{t}$, to minimize it. The choice of probability distortions is not unconstrained but rather subject to a penalty cost.

The lenders' preferences over consumption plans $\mathbf{c}^{L}$ after any history any $(t,w^{t})$ can be represented by the following specification
{\small{\begin{align}\label{eqn:BE_MA_3}
 U_{t}(\mathbf{c}^{L};w^{t}) =  c^{L}_{t}(w^{t}) + \gamma \hspace{-0.01in}\min_{m_{t+1}(\cdot | w^{t}) \in \mathcal{M}} \left\{ E_{Y} \left[ m_{t+1}(Y_{t+1}|w^{t}) \mathcal{U}_{t+1}(\mathbf{c}^{L};w^{t},Y_{t+1}) \mid y_{t} \right] +\theta
    \mathcal{E}[m_{t+1}(\cdot|w^{t})](y_{t}) \right\},
\end{align}}}
where $\gamma \in (0,1)$ is the discount factor, the parameter $\theta \in (\underline{\theta },+\infty ]$ is a penalty parameter
that measures the degree of concern about model misspecification\footnote{The lower bound $\underline{\theta }$ is a breakdown value below which the minimization problem is not well-behaved; see \cite{HS_book} for details.},  and the mapping $\mathcal{E} : \mathcal{M} \rightarrow L^{\infty}(\mathbb{Y})$ is the \emph{conditional relative entropy}, defined as
\begin{align} \label{eqn:entropy}
  \mathcal{E}[\lambda](y) \equiv E_{Y} \left[ \lambda(Y')  \log
\lambda (Y')   \mid y \right]
\end{align}
for any $\lambda \in \mathcal{M}$ and $y \in \mathbb{Y}$. Finally, $\mathcal{U}_{t+1}(\mathbf{c}^{L};w^{t},y_{t+1})$ is the expected value of {\small{$U_{t}(\mathbf{c}^{L};w^{t},y_{t+1},X_{t+1})$}}, conditioned on $y_{t+1}$, but before the realization of $X_{t+1}$, i.e.
\begin{equation}
\mathcal{U}_{t+1}(\mathbf{c}^{L};w^{t},y_{t+1}) \equiv E_{X}\left[ U_{t+1}(\mathbf{c}^{L};w^{t},y_{t+1},X_{t+1})  \right], \label{eqn:ExValue}
\end{equation}
and $U_{t}(\mathbf{c}^{L};w^{t})$ is the present value expected utility at time $t$, given that the previous history is given by $w^{t}$ and the agent follows a consumption plan $\mathbf{c}^{L}$. By looking at expression (\ref{eqn:BE_MA_3}) we see that the probability distortion $m_{t+1}$ pre-multiplies the expected continuation value before the realization of $X_{t+1}$, i.e. $\mathcal{U}_{t+1}(\mathbf{c}^{L};w^{t},y_{t+1})$, in line with our measurability assumption. For the sequential formulation of the lenders' lifetime utility and the derivation of the recursion (\ref{eqn:BE_MA_3})-(\ref{eqn:ExValue}), see Section \ref{app:PO_MA} in the Supplementary Material.

For any given history $w^{t}$, $\mathcal{E}[m_{t+1}(\cdot|w^t)](y_{t})$ measures the discrepancy of the distorted conditional probability, $\widetilde{P}_{Y,t}(\cdot \mid w^{t})$, with respect to the approximating conditional probability $P_{Y^{\prime}|Y}(\cdot |y_{t})$.
 Through this entropy term, the minimizing agent is penalized
whenever she chooses distorted probabilities that differ from the
approximating model. The higher the value of $\theta $, the more the minimizing agent is penalized. In the extreme case of $\theta =+\infty $, there are no concerns about model misspecification.

%

\subsection{Lenders' Problem}
\label{sec:LENDERS_T1}

As it will become clear below, for the recursive equilibrium in our particular environment, the lifetime utility in the previous section becomes $W_{R}(w_{t},B_{t},b_{t})$ or $W_{A}(y_{t})$.\footnote{As we will see, $x_{t}$ is not a state variable for the lender's problem in financial autarky due to its i.i.d. nature, and the fact that there is no decision making during that stage.} Here, $W_{R}(w_{t},B_{t},b_{t})$ is the equilibrium value (in lifetime utility) of an individual lender with access to financial markets, given the state of the economy $(w_{t},B_{t},b_{t})$. $W_{A}(y_{t})$  is analogously defined, but when the borrowing economy has no access to financial markets.


Since lenders are atomistic, each individual lender takes as given the price function and the aggregate debt $B_{t}$. The lender has a \emph{perceived} law of motion for this variable, which only in
equilibrium will be required to coincide with the actual one.\footnote{Remember that we denote $%
b_{t}$ as the individual lender's debt, while $B_{t}$ refers to the representative lender's debt.}

When lender and borrower can engage in a new financial relationship, the lender's min-max problem at state $(w_{t},B_{t},b_{t})$, is given by: {\small{
\begin{align}\label{eqn:BE_len}
  W_{R}(w_{t},B_{t},b_{t})& = \min_{m_{R} \in \mathcal{M}} \max_{c^{L}_{t},b_{t+1}}\left\{c^{L}_{t}  + \theta \gamma \mathcal{E}[m_{R}](y_{t}) + \gamma  E_{Y} \left[ m_{R}(Y_{t+1}) \mathcal{W}(Y_{t+1},X_{t+1},B_{t+1},b_{t+1}) |y_{t} \right] \right\} \notag \\
s.t.~c^{L}_{t}& =\overline{z}+q(y_{t},B_{t+1})(b_{t+1}-(1- \lambda) b_{t}) -(\lambda + (1-\lambda)\psi)b_{t}~and~B_{t+1}=\Gamma(w_{t},B_{t}),
\end{align}}}
where for all $y_{t+1} \in \mathbb{Y}$ the continuation value $\mathcal{W}(y_{t+1},B_{t+1},b_{t+1})$ is given by $\mathcal{W}(y_{t+1},B_{t+1},b_{t+1}) \equiv E_{X}\left[ W(y_{t+1},X_{t+1},B_{t+1},b_{t+1}) \right]$, where $W(w_{t+1},B_{t+1},b_{t+1})\equiv \delta
(w_{t+1},B_{t+1})W_{R}(w_{t+1},B_{t+1},b_{t+1})+(1-\delta(w_{t+1},B_{t+1}))W_{A}(y_{t+1})$ is the value of the lender when the borrower is given the option to default at state $(w_{t+1},B_{t+1},b_{t+1})$, and  $\Gamma: \mathbb{W} \times \mathbb{B}\rightarrow \mathbb{B}$ is the perceived law of motion of the individual lender for the debt holdings of the borrower, $B_{t+1}$. Observe that the optimal choice of $m_{R}$, is a mapping from $(w_{t},B_{t},b_{t}) \in \mathbb{W} \times \mathbb{B}^{2}$ to $\mathcal{M}$.

From equation (\ref{eqn:BE_len}) we note that lenders receive every period a non-stochastic endowment given by $\overline{z}$. Since the per-period utility is linear in consumption, the level of $\overline{z}$ does not affect  the equilibrium bond prices, bond holdings, and default strategies in our original economy; see Lemma \ref{lem:lenders_size}.

In financial autarky, as with the borrower, the lender has no decision to make. The  lender's autarky value at state $(y_{t})$, is thus given by
\begin{equation*}
W_{A}(y_{t})= \min_{m_{A} \in \mathcal{M}}  \left\{  \bar{z}+  \theta \gamma \mathcal{E}[m_{A}](y_{t}) + \gamma  E_{Y} \left[ m_{A}(Y_{t+1}) \left( (1-\pi
)W_{A}(Y_{t+1})+\pi \mathcal{W}(Y_{t+1},0,0) \right) \mid y_{t} \right]   \right\},
\end{equation*}
where $\pi$ is the re-entry probability to financial markets. Note that the optimal choice, $m_{A}$, is a mapping from $\mathbb{Y}$ to $\mathcal{M}$. In contrast with the borrower's case, no output loss is assumed for the lender during financial autarky.

\section{Recursive Equilibrium}

\label{sec:EQ_T1}

As is standard in the quantitative sovereign default models, we are interested in a recursive equilibrium in which all agents choose
sequentially.

\begin{definition}
	A collection of policy functions $\{c,c^{L},B,b,m_{R},m_{A},\delta \}$\ is given
	by mappings for consumption $c:\mathbb{W}\times \mathbb{B}\rightarrow \mathbb{R}_{+}$ and $c^{L}:\mathbb{W}\times \mathbb{B}%
	^{2}\rightarrow \mathbb{R}_{+}$, bond holdings $B:\mathbb{W}\times
	\mathbb{B}\rightarrow \mathbb{B}$\ and $b:\mathbb{W}\times \mathbb{B}%
	^{2}\rightarrow \mathbb{B}$\ for borrower and individual lender,
	respectively; and, probability distortions $m_{R}: \mathbb{W} \times \mathbb{B}^{2} \rightarrow \mathcal{M}$, $m_{A}: \mathbb{Y} \rightarrow \mathcal{M}$ and default decisions, $\delta : \mathbb{W} \times \mathbb{B}\rightarrow \{0,1\}.$	A collection of value functions $\{V_{R},V_{A},W_{R},W_{A}\}$%
	\ is given by mappings $V_{R}:\mathbb{W}\times \mathbb{B}\rightarrow
	\mathbb{R}$, $V_{A}:\mathbb{W}\rightarrow \mathbb{R}$, $W_{R}:%
	\mathbb{W}\times \mathbb{B}^{2}\rightarrow \mathbb{R}$, $W_{A}:%
	\mathbb{Y}\rightarrow \mathbb{R}.$
\end{definition}

%

\begin{definition}\label{def:equil}
A recursive equilibrium for our economy is a collection of policy functions $\{c^{\ast},c^{L, \ast},$ $B^{\ast },b^{\ast},m_{R}^{\ast},m_{A}^{\ast},\delta ^{\ast }\}$, a
collection of value functions $\{V_{R}^{\ast},V_{A}^{\ast},W_{R}^{\ast},W_{A}^{\ast}\}$, a perceived law of motion for the borrower's bond holdings, and a price schedule such that:

\begin{enumerate}
\item Policy
functions, probability distortions, and value functions solve the borrower and individual lender's optimization problems.

\item For all $(w,B) \in \mathbb{W} \times \mathbb{B}$, bond prices $q(y,B^{\ast }(w,B))$ clear the financial markets, i.e., $B^{\ast }(w,B)=b^{\ast }(w,B,B)$.

\item The actual and perceived laws of motion for debt holdings coincide,
i.e., $B^{\ast }(w,B)=\Gamma (w,B)$, for all $(w,B) \in \mathbb{W} \times \mathbb{B}$.
\end{enumerate}
\end{definition}


After imposing the market clearing condition above, vector $(w_{t},B_{t})$ is sufficient to describe the state variables for any agent in this economy. Hence, from here on, we consider $(w_{t},B_{t})$ as the state vector, common to the borrower and the individual lenders.

\subsection{Equilibrium Bond Prices and Probability Distortions}
\label{sec:EQ_PRICES}

In our competitive sovereign debt market, uncertainty-averse lenders make zero profits in expectation given their subjective beliefs. \footnote{In \cite{ARELLANO_AER08} competitive risk-neutral lenders are indifferent between any individual debt holdings level $b_{t+1}$. In our environment this is not true anymore. Taking $q(Y_{t+1},B^{\ast}(W_{t+1},B_{t+1}))$ and the borrower's strategies as given, lenders solve a convex optimization problem with a strictly concave objective function and hence there is a unique interior solution for individual debt holdings.} Hence, for an endowment level $y_{t}$ and a loan size $B_{t+1}$, the bond price function satisfies for all $(y_{t},B_{t+1})$,
\begin{align} \label{eqn:pricefn}
q(y_{t},B_{t+1})= \gamma E_{Y} \left[\chi(Y_{t+1};B_{t+1}) \mathbf{m}^{\ast}(Y_{t+1};y_{t},B_{t+1})\Big|y_{t}\right],
\end{align}
where $\chi:\mathbb{Y}\times\mathbb{B}\rightarrow \mathbb{R_{+}}$ denotes the payoff of the long-term bond, given by
\begin{align*}
\chi(y_{t+1};B_{t+1}) \equiv E_{X}\left[\Big(\lambda + (1-\lambda)\big(\psi+q(y_{t+1},B^{\ast}(w_{t+1},B_{t+1}))\big)\Big) \delta^{\ast}(w_{t+1},B_{t+1})\right]
\end{align*}
and $\mathbf{m}^{\ast}:\mathbb{Y}^{2}\times\mathbb{B}\rightarrow \mathbb{R_{+}}$ is given by
\begin{align*}
  \mathbf{m}^{\ast}(y_{t+1};y_{t},B_{t+1}) \equiv \frac{\exp \left\{ -\frac{\mathcal{W}^{\ast}(y_{t+1},B_{t+1},B_{t+1})}{\theta }\right\} }{E_{Y} \left[\exp \left\{ - \frac{ \mathcal{W}^{\ast}(Y_{t+1},B_{t+1},B_{t+1})}{\theta }\right\}\big|y_{t} \right]}.
\end{align*}
The function $\mathbf{m}^{\ast}$ is essentially the reaction function for the probability distortions, which is consistent with the FOCs in the minimization problem (\ref{eqn:BE_len}) and the market clearing condition for debt.\footnote{Observe that, by construction, $\mathbf{m}^{\ast}(y_{t+1};y_{t},B^{\ast}(w_{t},B_{t})) = m^{\ast}_{R}(y_{t+1};w_{t},B_{t})$. While $m^{\ast}_{R}$ are the optimal probability distortions along the equilibrium path, commonly computed in the robust control literature for atomistic agents, the reaction function $\mathbf{m}^{\ast}$ is a necessary object of interest in this environment to evaluate alternative debt choices for the borrower. }


Given the state of the economy next period, if defaults occurs, the payoff of the bond is zero. Otherwise, a fraction $\lambda$ of the bond matures while the remaining $(1-\lambda)$ pays off a coupon $\psi$ and keeps a market value of $q(Y_{t+1},B^{\ast}(W_{t+1},B_{t+1}))$.


In the absence of fears about model uncertainty, i.e. $%
\theta =+\infty$, the probability distortion vanishes, i.e. $\mathbf{m}^{\ast }=1$, that means that lenders' beliefs coincide with the approximating distribution $P_{Y'|Y}$, and hence the price function (\ref{eqn:pricefn}) is the same as in the rational expectations environment of \cite{Chatterjee_2010}.

\begin{remark}[One Period Bonds]
	In the case with one-period bonds, that is, $\lambda=1$ and setting $X_{t}=0$, the pricing equation \ref{eqn:pricefn} collapses to
	\begin{align} \label{eqn:pricefn_op}
	q(y_{t},B_{t+1})= \gamma E_{Y} \left[  \delta^{\ast}(Y_{t+1},B_{t+1}) \mathbf{m}^{\ast}(Y_{t+1};y_{t},B_{t+1})\Big|y_{t}\right].
	\end{align}
	For $\theta = + \infty$ the expression coincides with that of \cite{ARELLANO_AER08}. \footnote{We refer the reader to our working paper version for a more thorough discussion and result for the one period bond case with $\theta < \infty$.}
\end{remark}

Under model uncertainty, the lender in this economy distrusts the conditional probability $P_{Y'|Y}$ and wants to guard
himself against a worst-case distorted distribution for $y_{t+1}$, given by $\mathbf{m}^{\ast }(\cdot;w_{t},B^{\ast}(y_{t},B_{t}))P_{Y^{\prime}|Y}(\cdot|y_{t})$. The fictitious minimizing agent will be selecting this worst-case density by slanting probabilities towards the states associated with low continuation utility for the lender. In the presence of default risk, the states associated with lowest utility coincide with the states in which the borrower defaults and therefore the lender receives no repayment.
In addition, in this economy with long-term debt, upon repayment, the payoff responds to variations in the next-period bond price. Hence, states in which the latter is lower will be associated with relatively higher probability distortions.

\begin{figure}[htp]
\centering
\includegraphics[height=3in,width=5.5in]{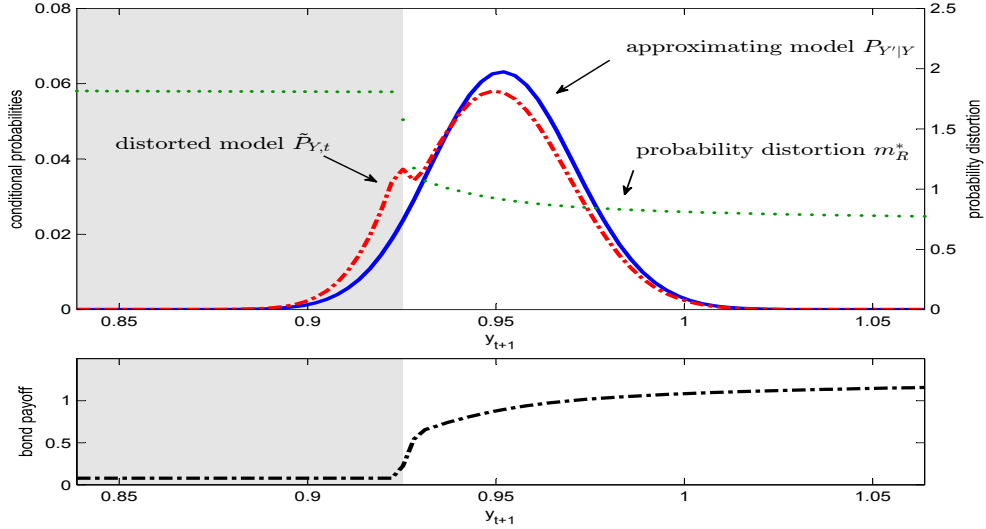}
\caption{{\footnotesize{Approximating and distorted densities.}}}
\label{fig:2_densities}
\end{figure}

Figure \ref{fig:2_densities} illustrates the optimal distorting of the probability of next period realization of $Y_{t+1}$, given current state $(w_{t},B_{t})$ with access to financial markets.\footnote{For illustrative purposes, a low endowment $y_{t}$ and low bond holdings $B_{t}$, or equivalently high debt level, were suitably chosen to have considerable default risk under the approximating density. The current endowment level $y_{t}$ corresponds to half a standard deviation below its unconditional mean, and the bond holdings $B_{t}$ are set to the median of its unconditional distribution in the simulations. Also, current $x_{t}$ was set to zero.} $B_{t+1}$ is computed using the optimal debt policy, i.e. $B_{t+1}=B^{\ast}(w_{t},B_{t})$. In the top panel of this figure we plot the conditional approximating density and the distorted density for $y_{t+1}$, as well as its corresponding probability distortion $m_{R}^{\ast}$ (with $ m^{\ast}_{R}(y_{t+1};w_{t},B_{t}) = \mathbf{m}^{\ast}(y_{t+1};y_{t},B^{\ast}(w_{t},B_{t})) $). The bottom panel plot depicts the expected payoff of the bond $\chi(Y_{t+1};B_{t+1})$ at $t+1$ for each value of $y_{t+1}$, before the realization of $X_{t+1}$. Note that this expected payoff is continuous in $Y_{t+1}$ as a result of the smoothing of the conditional default probability by the output shock $X_{t+1}$. The shaded area in both panels corresponds to the range of values for the realization of $y_{t+1}$ in which the borrower defaults with probability equal or higher than 50 percent (note that the default decision at $t+1$ also depends on the realization of $X_{t+1}$).

In order to minimize lenders' expected utility, the minimizing agent places a non-smooth probability distortion $m_{R}^{\ast}(\cdot;w_{t},B_{t})$ over next-period realizations of $y_{t+1}$, with values strictly larger than 1 over the default interval, that drops dramatically below 1 as repayment becomes more certain. By doing so, the minimizing agent takes away probability mass from those states in which the borrower does not default,
and puts it in turn on those low realizations of $y_{t+1}$ in which default is optimal for the borrower. For this particular state vector $(y_{t},B_{t})$ in consideration, the conditional default probability under the approximating model is $9.3$ percent quarterly, while under the distorted one it is $16.2$ percent, almost twice as high. The kinked shape of $m_{R}^{\ast}$ follows from the kinked shape of the  lenders' utility value with respect to $y_{t+1}$, which in turn is due to the kinked shape in next-period payoff of the bond as function of $y_{t+1}$.\footnote{If there were no output shock, the payoff of the bond would be discontinuous due to the default contingency, and so would the probability distortion.} 

With long-term debt, additional probability distorting takes place over the repayment interval. Since the payoff of the bond remains state contingent due to its dependence on the next-period bond price, so will the lenders' utility. Consequently, states associated with relatively lower next-period prices will be assigned relatively higher weights. 

The tilting of the probabilities
by the minimizing agent generates an endogenous hump of the distorted density
over the interval of $y_{t+1}$ associated with default risk,
as observed in Figure \ref{fig:2_densities}. The bi-peaked form of the resulting distorted conditional density differs from the standard distortions in the robust control literature, in which it typically displays only a shift in the conditional mean from the approximating one.\footnote{See, for example, \cite{BHS} and \cite{AHS}.}$^{,}$\footnote{Table \ref{tab:MOM} in Section \ref{app:mom} of the Supplementary Material reports distortions in several moments of $y_{t}$ for our economy.}


Sovereign default events in our model can be interpreted as ``disaster
events'', which in our economy emerge endogenously from the borrower's decision making and the lack of enforceability of debt contracts. Fears about model misspecification in turn amplify their effect on both allocations and equilibrium
prices, as they increase the perceived likelihood in the mind of the lenders of these
rare events occurring. As a result, the model can be viewed as generating endogenously varying disaster risk.

\medskip

\begin{figure}[htp]
\centering
\includegraphics[height=3in,width=6in]{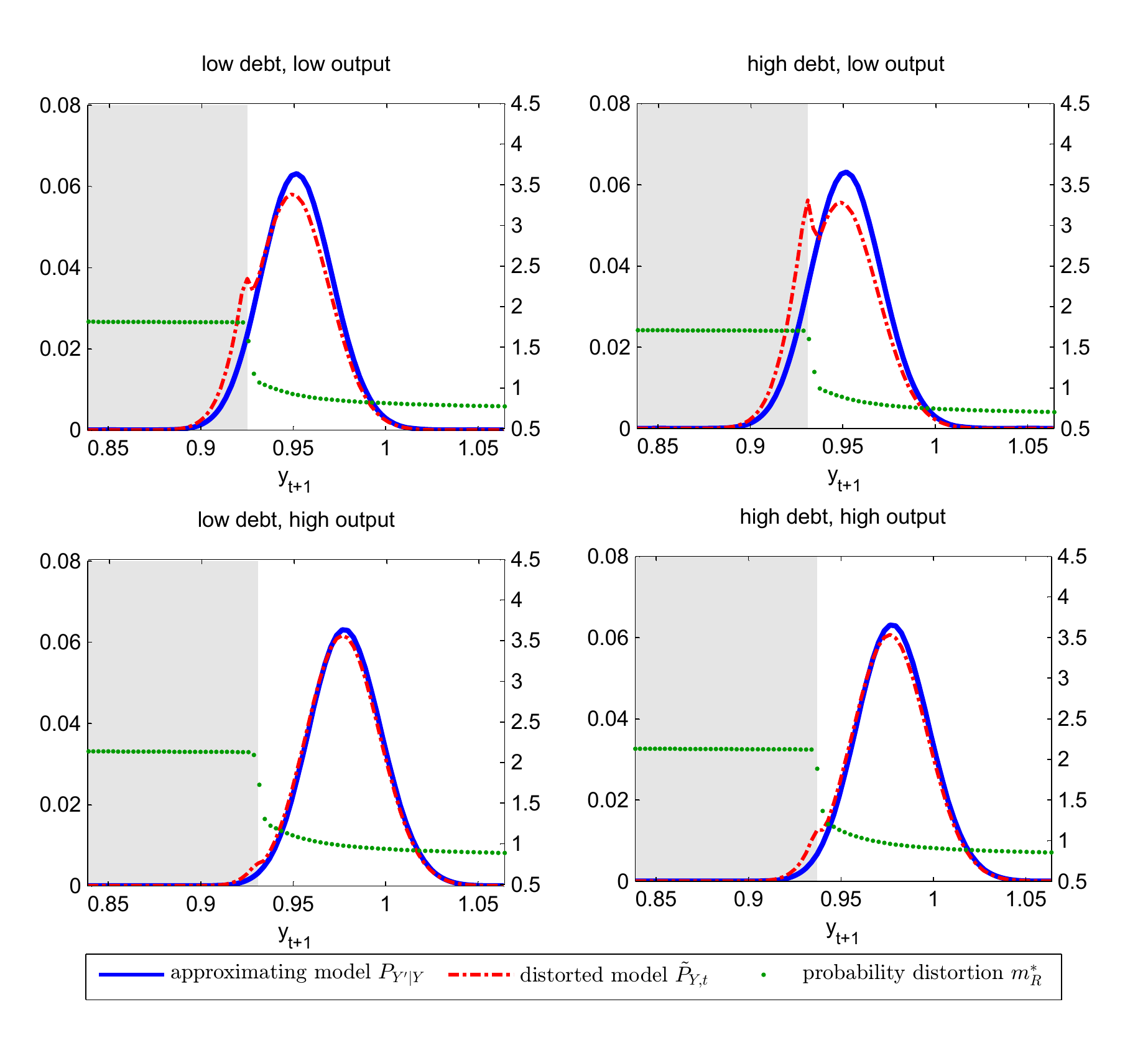}
\caption{{\footnotesize{Approximating and distorted densities for different state vectors $(y_{t},B_{t})$.}}}
\label{fig:3_densities_yb}
\end{figure}

\textbf{State-dependency of probability distortions.} In our economy, probability distortions are state-dependent and thereby typically time-varying. The default risk under the approximating density and the quantity of bonds carried over to next period, which the borrower can default on, affect the extent to which the minimizing agent distorts lenders' beliefs. 
Figure \ref{fig:3_densities_yb} shows the approximating and distorted density of next period $y_{t+1}$ for different combinations of current endowment and bond holdings, $(y_{t},B_{t})$.\footnote{Low and high endowment $y_{t+1}$ correspond to half and a quarter a standard deviation below the unconditional mean of $y_{t}$, respectively. The i.i.d. output shock $x_{t}$ is set again to zero. Also, low debt is given by the median of the debt unconditional distribution, and high debt corresponds to the 60th percentile.}

By comparing the two panels in the top row (or the bottom row), we can see how the probability distortion changes with the level of current debt. In this general equilibrium framework, we need to take into account the optimal debt response of the borrower for the current state of the economy. For the state vectors in consideration, the higher the current level of indebtness $B_{t}$, the more debt the borrower optimally chooses to carry into next period, $B_{t+1}$. To see how the \emph{perceived} probability of default next period varies, we check at how the default risk under the approximating model and the probability distortions change as current bond holdings $B_{t}$ increase. First, the interval of realizations of $y_{t+1}$ for which the borrower defaults is enlarged. The larger the quantity of bonds that the borrower has to repay at $t+1$, the greater the incentives it would have to not do it. Consequently, the default risk under the approximating model is higher. Also, those new states on which there is default with higher debt become now low-utility states for the lender, and hence probability distortions $m^{\ast}_{R}$ larger than 1 are assigned to them in the new, worst-case density.

Second, the change in levels of the probability distortions may not be straightforward. On the one hand, since for these cases, more bond holdings $B_{t+1}$ are carried into next period, more is at stake for the lender, as the potential losses in the event of default are larger. Hence, the probability mass on the default states would be even higher than before. One the other hand, higher $B_{t+1}$ also means higher default risk in the future, which would also depress next-period bond prices and thereby the payoff of the bond over the repayment interval. Since the optimal probability distortions are assigned on the basis of the relative payoff in each state, they may be higher or lower than before. While probability distortions (over the default interval) turn smaller in the top panels of Figure \ref{fig:3_densities_yb} as debt increases, the opposite occurs in the bottom ones.

By comparing the two panels in the left-side column (or the right-side column) we can see how the probability distortion changes with the level of current endowment. Due to the persistence of the stochastic process for $(y_{t})_{t}$, the lower the current endowment, $y_{t}$, the lower the conditional mean of next period's endowment, $y_{t+1}$ of the approximating density. For the state vectors considered here, the agent gets relatively more indebted as current endowment $y_{t}$ rises. This follows from the fact that output costs of default are increasing in the endowment $y_{t}$. The higher $y_{t}$, the more severely the borrower is punished if it defaults. As the incentives to default are smaller, the returns are lower, or equivalently the bond prices demanded by the lenders are higher, for the same levels of debt. Facing relatively cheaper debt, the borrower responds by borrowing more. In this way, more bond holdings $B_{t+1}$ widens the intervals of $y_{t+1}$-realizations for which default is optimal. At the same time, probability distortions over the new default interval become relatively larger for similar reasons as discussed previously when debt $B_{t}$ rises. In these cases, the \emph{perceived} probability distortions, however, decrease due to the rightward shift of the conditional mean of $y_{t+1}$, as endowment $y_{t}$ increases. 


\medskip

\textbf{Comparison to CRRA and Epstein-Zin Utilities.} A natural question is whether risk aversion on the lenders' side with time separable preferences could generate a stochastic discount factor, negatively correlated with default decisions $\delta^{\ast}_{t+1}$, that could help account for low bond prices, while preserving the default frequency at historical low levels. We explore this in Section \ref{app:RA} of the Supplementary Material. Our findings indicate that in our calibrated economy with CRRA separable preferences for the lender plausible degrees of risk aversion on the lenders' side are \emph{not} sufficient to generate high bond returns; to some extent, this is analogous to the equity premium puzzle result studied in \cite{Mehra_Prescott}. See Table \ref{tab:RA} for details. Our results are also consistent with the findings by \cite{Lizarazo} and \cite{Verdelhan}.

Even if sufficiently high values of risk aversion could eventually recover the high spreads shown in the data, doing so, however, would  lower the risk-free rate to levels far below those exhibited in the data, in line with \cite{Weil} risk-free rate puzzle.\footnote{Note that the stochastic process assumed for $Y_{t}$ is stationary. If we add a positive trend, the risk-free rate would be rising, rather than decreasing, as the lenders' coefficient of risk aversion goes up.}

In our environment with model uncertainty, however, the extent to which lenders are uncertainty-averse does not affect the equilibrium gross risk-free rate, given by the reciprocal of $\gamma$, as their period utility function is linear in consumption.\footnote{In our model with linear lenders' per-period utility, equilibrium prices depend exclusively on economic fundamentals of the borrowing economy and the lenders' preference for robustness. It is noteworthy to remark that adding curvature on the per-period utility will, in general, lead to equilibrium prices that also depend on international lenders' characteristics such as their total wealth and investment flows, or more generally, on global macroeconomic factors, in line with the empirical findings by \cite{Longstaff}. This seems to be an interesting extension to pursue in future research.}

In a wide class of environments, the utility recursion with multiplier preferences can be reinterpreted in terms of Epstein-Zin utility formulation.\footnote{We note that \cite{EZ-ECMA89} accommodates per-period payoff specifications that go beyond the log case.} In such a case, the typical probability distortion through which the agent's uncertainty aversion is manifested would take the form of a risk-sensitive adjustment used to evaluate future streams of consumption. In our framework, this apparent observational equivalence, however, does not apply, because the lender contemplates perturbations only to the probability model governing the evolution of the borrower's endowment, and not to the probability distribution of reentry to financial markets, which is assumed to be fully trusted. In our setup, since re-entry occurs with zero debt and it does not directly affect prices, we expect that perturbations on the probability of re-entry will not have a significant effect on the quantities of interest.

%
%

\medskip

\textbf{Irrelevance of lenders' wealth.} We conclude this section by showing that the size of lenders' non-stochastic endowment is irrelevant for equilibrium bond prices and the borrower's allocations.
\begin{lemma}\label{lem:lenders_size}
Consider an arbitrary recursive equilibrium with lenders' non-stochastic endowment given by $z$. Then, for any other non-stochastic endowment $\hat{y}^{L} \ne z$, there exists a recursive equilibrium with identical bond prices and borrower's allocations.
\end{lemma}

The proof and formal statement are deferred to the Appendix \ref{sec:yL}.\footnote{In the Appendix we actually prove a more general result, that allows for stochastic endowment for the lenders.} This result has important implications for our calibrations, because there is no need to identify who the lenders are in the data, and, in particular, to find a good proxy of their income relatively to the borrower's endowment.

Finally, we note that by similar calculations it can be shown that equilibrium bond prices or the borrower's allocation remain unchanged even if lenders were allowed to borrow or save at a given gross risk-free rate in positive net supply in credit markets, e.g. investing in U.S. Treasury bills.

\section{Quantitative Analysis}
\label{sec:quant}
In this section we analyze the quantitative implications of our model for Argentina. To do so, we specify our choices for functional forms and calibrate some parameter values to match key moments in the data for the Argentinean economy. The period considered spans from the first quarter of 1993, when Argentina regained access to financial market with the Brady Plan agreement after its 1982 default, to the last quarter of 2001, when Argentina defaulted again on its external debt.

\subsection{Calibration}
\label{sec:cali}

For the quantitative analysis, we consider the following functional forms. The period utility function for the borrower is assumed to have the CRRA form, i.e., $u(c)=\frac{c^{1-\sigma}}{1-\sigma}$ where $\sigma$ is the coefficient of relative risk aversion.

We assume that the endowment of the borrower follows a log-normal AR(1) process, $\log Y_{t+1}=\rho \log Y_{t}+\sigma_{\varepsilon }
\varepsilon _{t+1}$, where the shock $\varepsilon _{t+1}\sim \mathcal{N}(0,1)$. As shown in Lemma \ref{lem:lenders_size}, the (non-stochastic) lenders' endowment does not affect the equilibrium bond prices and borrower's allocations, allowing us to circumvent the subtle challenge of providing a good proxy for lenders' consumption or income. We therefore set the lenders' logged endowment, denoted by $\log(\overline{z})$, to 1.

Following \cite{Chatterjee_2010}, we consider the specification for output costs
\begin{equation} \label{eqn:ouputcosts}
\phi(y)=\max \left\{0,\kappa_{1}y+\kappa_{2}y^{2}\right\},
\end{equation}
with $\kappa_{2} > 0 $. As explained later, our calibrated output costs play a key role
in generating desired business cycle features for emerging economies, in particular the volatility
of bond spreads, in the context of lenders' model uncertainty. 
In Table \ref{tab:cali} we present the parameter values for the calibration of our benchmark model.

The coefficient of relative risk aversion for the borrower $\sigma$ is set to $2$,
which is standard in the sovereign default literature. The re-entry
probability $\pi $ is set to $0.0385$, implying an average
period of $6.5$ years of financial exclusion and consistent with \cite{Benjamin_Wright} estimates.\footnote{\cite{Pitchford_Wright} report
an average $6.5$-year delay in debt restructuring after 1976.}

We estimate the parameters $\rho$ and $\sigma_{\epsilon}$ for the log-normal AR(1) process
for the endowment of the borrower, using output data for Argentina for the period from 1993:Q1 to 2001:Q3.\footnote{We exclude the last quarter of 2001 since the default announcement by President Rodriguez Sa\'{a} took place on December 23, 2001.}$^{,}$\footnote{Time series at a quarterly frequency for output, consumption, and net exports for Argentina are taken from the Ministry of Finance (MECON). All these series are seasonally adjusted, in logs, and filtered using a linear trend. Net exports are computed as a percentage of output.}

The (one-period) risk-free rate $r^{f}$ in the model is $1$ percent, which is approximately the average quarterly interest rate of a three-month U.S. Treasury bill for the period in consideration. The lenders' discount factor $\gamma$ is set equal to the reciprocal of the gross risk-free rate $1+r^{f}$. \footnote{Lenders can be allowed to trade a zero net supply risk-less claim to one unit of consumption next period. Since all lenders are identical, no trade in such a claim takes place in equilibrium, where $1 + r^{f} = 1/\gamma$.} The parameters governing the payoff structure of the long-term bond, $\lambda$ and $\psi$, are chosen to replicate a median debt maturity of $5$ years and a coupon rate of $12$ percent.

Bond spreads are computed as the difference between annualized bond returns and the U.S. Treasury bill rate. The quarterly time series on interest rate for sovereign debt for Argentina is taken from \cite{Neumeyer_Perri}. To calculate the yield of the long-term bond, we use the internal rate of return.\footnote{The internal rate of return of a bond, denoted by $r(y_{t},B_{t+1})$, is determined by the pricing equation:
\begin{equation*}
q(y_{t},B_{t+1})=\frac{\lambda + (1-\lambda)\psi}{\lambda + r(y_{t},B_{t+1})}.
\end{equation*}
}

We calibrate the parameters $\beta$, $\kappa_{1}$, $\kappa_{2}$, and $\theta$ in our model to match key moments for the Argentinean economy. We set the borrower's discount factor $\beta$ to target an annual frequency of default of $3$ percent. The calibrated value for $\beta$ is $0.9627$, which is relatively large within the sovereign default literature.\footnote{For example, \cite{Yue} uses a discount factor of $0.74$ and \cite{AG_1} use $0.80$.}

We select the output cost parameters $\kappa_{1}$ and $\kappa_{2}$ to match the average debt level of 46 percent of GDP for Argentina and the spreads volatility of 4.58 percent.\footnote{The external government debt to output ratio of 46 percent for Argentina is taken from the National Institute of Statistics and Census (INDEC) for the period from 1994:Q4 to 2001:Q4.}

 Regarding the degree of model uncertainty in our economy, we take the following strategy: we first set the
 penalty parameter $\theta$ to match the average bond spreads of $8.15$ percent observed in the data for Argentina.
 As pointed out by \cite{BHS}, the value of $\theta$ is itself not necessarily informative of the amount of distortion in lenders' perceptions about the evolution of $y$; its impact on probability distortions is context-specific.\footnote{See \cite{BHS} for a simple example with a random walk model and a trend stationary model for log consumption.}

To better interpret our results, we provide another statistic, the detection error probabilities (DEP), commonly used
 in the robust control literature.\footnote{See \cite{AHS}, \cite{Maenhout}, \cite{BHS}, \cite{Bidder_Smith}, and \cite{LNY}, for example.}
 For a more thorough discussion, details on its computation and an alternative measure, see Section \ref{sec:dep}. The lower the value of DEP, the more pronounced is the discrepancy between these two models. If they are basically identical, they are indistinguishable and hence the DEP is $0.50$. In contrast, if the two models are perfectly distinguishable from each other, the DEP is $0$. \cite{BHS} suggest $20$ percent as a reasonable threshold, in line with a 20-percent Type I error in statistics. In our model the DEP implied by our calibrated $\theta$ is only 31 percent, which implies that around one third of the time the detection test indicates the wrong model. This value is therefore quite conservative, suggesting that only a modest amount of model uncertainty is sufficient to explain the high average bond spreads observed in the data.

\begin{table}[ht]
  \centering
{\footnotesize{  \begin{tabular}{lll|l}
    \hline\hline
&  & Parameter & Value \\ \hline
Borrower  & Risk aversion & \multicolumn{1}{c|}{$\sigma$} & \multicolumn{1}{|c}{$%
2 $} \\
& Time discount factor & \multicolumn{1}{c|}{$\beta$} &
\multicolumn{1}{|c}{$0.9627$} \\
& Probability of reentry & \multicolumn{1}{c|}{$\pi $} & \multicolumn{1}{|c}{%
$0.0385$} \\
& Output cost parameter & \multicolumn{1}{c|}{$\kappa_{1} $} & \multicolumn{1}{|c}{$-0.255$}
\\
& Output cost parameter & \multicolumn{1}{c|}{$\kappa_{2} $} &
\multicolumn{1}{|c}{$0.296$} \\
& AR(1) coefficient for $y_{t}$ & \multicolumn{1}{c|}{$\rho $} &
\multicolumn{1}{|c}{$0.9484$} \\
& Std. deviation of $\varepsilon _{t}$ & \multicolumn{1}{c|}{$\sigma
_{\varepsilon }$} & \multicolumn{1}{|c}{$0.02$} \\
& Std. deviation of $x _{t}$ & \multicolumn{1}{c|}{$\sigma
_{x}$} & \multicolumn{1}{|c}{$0.03$} \\
&  &  &  \\
Lender & Robustness parameter & \multicolumn{1}{c|}{$\theta $} &
\multicolumn{1}{|c}{$0.619$} \\
& Constant for $z$ & \multicolumn{1}{c|}{$\log(\overline{z})$} &
\multicolumn{1}{|c}{$1.00$} \\
&  &  &  \\
Bond  & Risk-free rate & \multicolumn{1}{c|}{$r^{f}$} &
\multicolumn{1}{|c}{$0.01$} \\
& Decay rate & \multicolumn{1}{c|}{$\lambda$} &
\multicolumn{1}{|c}{$0.05$} \\
& Coupon & \multicolumn{1}{c|}{$\psi$} &
\multicolumn{1}{|c}{$0.03$} \\ \hline\hline
  \end{tabular}
  \caption{Parameter Values}
  \label{tab:cali}}}
\end{table}

\bigskip

\textbf{Computational algorithm. } The model is solved numerically using value function iteration. To that end, we apply the discrete state space (DSS) technique. The endowment space for $y_{t}$ is discretized into 200 points and the stochastic process is
approximated to a Markov chain, using \cite{Tauchen} quadrature-based method.\footnote{For bond holdings, we use 580 gridpoints to solve the model and no interpolation. Also, the distribution for $x_{t}$ is truncated between $[-2\sigma_{x},2\sigma_{x}]$.}

When solving the model using the DSS technique, we may encounter lack of convergence problems; see \cite{Chatterjee_2010} for details. To avoid that for long-term debt, we introduce the i.i.d. continuous output shock $X_{t}$.

To compute the business cycle statistics, we run $2,000$ Monte Carlo (MC)
simulations of the model with $4,000$ periods each.\footnote{To avoid dependence on initial conditions, we pick only the last 2,000 periods from each simulation. The unconditional default frequency is computed as the sample mean of the number of default events in the simulations.} Similarly to \cite{ARELLANO_AER08}, to replicate the period for Argentina from 1993:Q1 to 2001:Q3, we consider 1,000 sub-samples of 35 periods with access to financial markets, followed by a default event.\footnote{Because Argentina exited financial autarky with the Brady bonds while in our model it does so with no debt obligations, we also impose no reentry in the previous four quarters (1 year) of each candidate sub-sample.} We then compute the
mean statistics and the 90-percent confidence intervals, across MC simulations, for these subsamples.

\bigskip
\textbf{Output costs and implications. } The choice of \cite{Chatterjee_2010} specification for the output costs of default given by expression (\ref{eqn:ouputcosts}) is key for matching some business cycle moments.

Similar to their calibration, we have $\kappa_{1}<0$, which implies that there are no output costs for realizations $y < \kappa_{2}/\kappa_{1}$, and the output costs as a fraction of output increase with $y$ for $y > \kappa_{2}/\kappa_{1}$. In this sense, our output costs are similar to those in \cite{ARELLANO_AER08}, both of which have significant implications for the dynamics of debt and default
events in the model. As explained before, when output is high, there is typically less default risk, bond returns are low and there is more borrowing. For low levels of output, the costs of default are lower, hence, the default risk is higher, and so are the bond returns. If the borrower is hit by a sequence sufficiently long of bad output realizations, it eventually finds it optimal to declare default.

As noted by \cite{Chatterjee_2010}, this functional form for output costs has an important advantage over those of \cite{ARELLANO_AER08} for the volatility of bond spreads. In \cite{ARELLANO_AER08} output costs as a fraction of output vary significantly with output, and so do the default incentives.\footnote{Quantitatively, in that specific framework with one-period bonds, spreads volatility can be considerably reduced when using very fine grids or alternative computational methods to solve the model numerically, as shown by \cite{Hatchondo_RED}. For long-term debt models, however, no comparison between solution methods has been driven.} Hence, the default probability is very sensitive to the endowment realizations $y$. In our model, the sensitivity of the distorted default probability is even higher. Beliefs' distortions play out in the same direction, by slanting probabilities even more towards the range of endowment realizations in which default occurs. As a result, not surprisingly, the variability of bond spreads rises significantly when we introduce doubts about model misspecification.

For this reason, instead of using the output cost structure from \cite{ARELLANO_AER08}, we consider the specification given by (\ref{eqn:ouputcosts}). In this case, the output loss as a proportion of output is less responsive to fluctuations of $y$. It therefore yields a lower sensitivity of default probabilities to $y$, reducing at the same time the spreads volatility.

\subsection{Simulation Results}
\label{sec:quan}

Table \ref{tab:ARE_US} reports the moments of our calibrated model and in the data. For comparison purposes, it also shows the corresponding moments for \cite{Chatterjee_2010}, denoted CE model, probably the best-performing long-term debt model in the literature. Apart from model uncertainty, a key feature along which our calibrated model and the CE model differ is the targeted default frequency (3 percent in th former vis-a-vis 6.6 percent in the latter). For this reason, we introduce a re-calibrated version of our model that targets a default frequency of 3 percent but without model uncertainty and denote it the Baseline model. While the theoretical CE model and our baseline model are identical, their calibrations differ along several other dimensions. In particular, the targeted debt-to-output ratios are different: 70 percent in the CE model while 46 percent in the baseline model.\footnote{As mentioned before, in our calibration we consider the external government debt-to-output ratio of 46 percent for Argentina taken from the INDEC, which is similar to the debt levels reported by \cite{ARELLANO_AER08}, \cite{Yue}, and \cite{Mendoza_Yue}. In contrast, \cite{Chatterjee_2010} use as debt the total long-term public and publicly guaranteed external debt, provided by the World Bank’s Global Development Finance Database (GDF), which totaled 70 percent of GDP. Besides targeting different moments in the data, a different parametrization for the AR(1) endowment process is considered, as well as different number of asset grid-points and sampling criterion are used.}

\begin{table}[ht]
	\centering
	{\small{  \begin{tabular}{l|c|ccc}
				\hline\hline
				Statistic & Data & CE Model & Baseline Model   & Our Model \\ \hline
				Mean$(r-r^{f})$ & $8.15$ & $8.15$ &  $5.01$   & $8.15$  \\
				Std.dev.$(r-r^{f})$ & $4.58$ & $4.43$ &  $4.27$   & $4.62$ \\
				mean$(-b/y)$ & $46$ & $70$  & $42$ &   $44$ \\
				Std.dev.$(c)/$std.dev. $(y)$ & $0.87$  & $1.11$ &  $1.16$ & $1.23$  \\
				Std.dev.$(tb/y)$ & $1.21$ & $1.46$ & $0.89$  & $1.23$ \\
				Corr$(y,c)$ & $0.97$ & $0.99$ & $0.99$   & $0.98$ \\
				Corr$(y,r-r^{f})$ & $-0.72$ & $-0.65$  & $-0.78$   & $-0.75$ \\
				Corr$(y,tb/y)$ & $-0.77$ & $-0.44$  & $-0.80$   & $-0.68$ \\
				& \multicolumn{1}{|l|}{} &  &  &  \\
				Drop in y (around default) & $-6.4$ & $-4.5$ & $-3.9$ & $-5.6$ \\
				DEP& $NA$ & $NA$ &  $50.0$  &$31.3$\\
				& \multicolumn{1}{|l|}{} &  &  &  \\
				Default frequency (annually) & $3.00$ & $6.60$ & $3.00$   & $3.00$ \\ \hline \hline
			\end{tabular}
			\caption{Business Cycle Statistics for the data, the CE model, the Baseline model and our model.}
			\label{tab:ARE_US}}}
\end{table}

Overall, our model matches standard business cycle regularities of the Argentinean economy. More importantly, we can replicate salient features of the bond spreads dynamics. By introducing doubts about model misspecification, we can account for all the average bond spreads
observed in the data, as well as their volatility, matching at the same time the historical annual frequency of
default of $3$ percent and the average risk-free rate. 
An important contribution of our paper is that we only require quite limited amount of model uncertainty to do it. Indeed, we need on average smaller deviations of lenders' beliefs to explain the spread dynamics than those used in the equity premium literature.\footnote{To explain different asset-pricing puzzles, \cite{Maenhout}, \cite{Drechsler}, and \cite{Bidder_Smith} require a detection error probability in the range between $10$ and $12$ percent. \cite{BHS} needs even lower values to reach the \cite{Hansen_Jagannathan} bounds.}$^{,}$\footnote{It is worth noting that while we assume no recovery on defaulted debt in the model---which in equilibrium pushes up the bond returns---, there is room to increase the amount of model uncertainty (i.e., decrease $\theta$) within the plausible range, and thus we could still account for the bond spread average level if any mechanism of debt restructuring with subsequent haircuts is introduced.}

Notably, our model can explain the average bond spreads of $8.15$ percent in the data, which is roughly three percentage points higher than the $5.01$ percent obtained by the baseline model. 
  In our environment, risk-neutral lenders charge an additional uncertainty premium on bond holdings to get compensated for bearing the default risk under the worst-case density for output. In turn, their \emph{perceived} conditional probability of default next period ---while having access to financial markets--- is on average $2.2$ percent per quarter, while the actual one is only $0.9$ percent. Lenders' distorted beliefs about the evolution of the borrowing economy enable us to achieve the challenging goal of simultaneously matching the low sovereign default frequency and the high average level (and volatility) of excess returns on Argentinean bonds exhibited in the data. Additionally, our model can account for a strong countercyclicality of bond spreads.

As shown in Table \ref{tab:ARE_US}, \cite{Chatterjee_2010} (CE model) has also been able to match the
average bond spreads observed in the data. To our knowledge, only this paper and \cite{Hatchondo_Debt_Dilution} have been able to do that under rational expectations. 
These authors, however, reproduce the average high spreads for Argentina at the cost of roughly doubling the default frequency to $6.6$ percent annually. While it is difficult to determine what the true value for the default frequency is in the data, it seems to be consensus in the literature that it lies close to $3$ percent per year (see footnote 7). These papers and ours replicate this feature of the bond spreads in a general equilibrium framework. In contrast, \cite{ARELLANO_AER08}
and \cite{Arellano_WP}\footnote{See also \cite{Hatchondo_Debt_Dilution}.},
have been able to account for the bond spreads dynamics by assuming an ad hoc functional form
for the stochastic discount factor, which depends on the output shock to
the borrowing economy. Our paper can be seen as providing
microfoundations for such a functional form.\footnote{Indeed, the ad-hoc pricing kernels used in these studies can be reinterpreted as a probability distortion that alters the conditional mean but not the variance of the log-normal distribution of the endowment of the borrower. Section \ref{sec:kernel} in the Supplementary Material elaborates on this point.}


\begin{table}[ht]
	\centering
	{{  \begin{tabular}{l|c|cp{0.05cm}c}
				\hline\hline
				Statistic & Data & Baseline Model &  & Our Model \\ \hline
				$Q_{0.10}(r-r^{f})$ & $4.40$ & $2.12$ &  & $4.66$ \\
				$Q_{0.25}(r-r^{f})$ & $5.98$ & $2.62$ &   & $5.38$  \\
				$Q_{0.50}(r-r^{f})$ & $7.42$ & $3.57$ &  & $6.66$  \\
				$Q_{0.75}(r-r^{f})$ & $8.45$ & $5.55$ &  & $9.08$  \\
				$Q_{0.90}(r-r^{f})$ & $11.64$ & $9.65$ &  & $13.61$  \\       \hline\hline
			\end{tabular}
			\caption{Quantiles of spreads for our Model,
				the Data and the Baseline Model. $Q_{\alpha}(r-r^{f})$ denotes the $\alpha$-th quantile. }
			\label{tab:q_ARE_US}}}
\end{table}

In order to shed more light on the behavior of the spreads, we report in Table \ref{tab:q_ARE_US} different percentiles. In all cases, the average across MC simulations is very close to the one observed in the data. The baseline model, however, yields percentiles that are considerably below the values observed in the data. Finally, we note that the median is always lower than the average in the data and in the models, due to the presence of large peaks because of the default events. These results show how our model is able to match the average level, volatility and countercyclicality of spreads, while not distorting other relevant moments.

Also, in Section \ref{app:RA} of the Supplementary Material we provide simulations that show that the introduction of plausible degrees of risk aversion on the lenders'
side with time-separable preferences is insufficient to recover the high spreads observed in
the data. With constant relative risk aversion, as in \cite{Lizarazo},
matching high spreads calls for a very large risk aversion coefficient and
implausible risk-free rates.\footnote{\cite{Verdelhan} have studied the
setup with positive co-movement between lenders' consumption and output in
the emerging economy in addition to time-varying risk aversion on the
lenders' side. To generate endogenous time-varying risk aversion for
lenders, they endow them with \cite{Cochrane} preferences with external
habit formation. However, they find that even with these additional
components average bond spreads generated by the model are far below from
those in the data. They report average bond spreads of $%
4.27$ percent, for an annual default frequency of $3.11$ percent.}

Our model can also generate considerable levels of borrowing, consistent with levels observed in the data. High output costs of default jointly with a low probability of regaining access to financial markets imply a severe punishment to the borrower in case it defaults. Consequently, higher levels of indebtness can be sustained in our economy. Since the magnitude of output cost can be hard to gauge from the parameter values $\kappa_{1}$ and $\kappa_{2}$, we report the average output drop that the borrowing economy suffers in the periods of default announcements. We then compare this statistic with the actual contraction in Argentinean output observed in the data around the fourth quarter of $2001$, which reached -6.4 percent; in our model this number is -5.6 percent.\footnote{To be consistent with our model, the same linear trend from the estimation was employed when computing the actual drop of output in the data.}$^{,}$\footnote{The drop observed in $2002Q1$ was $7.3$ percent, slightly larger than in the previous quarter.} 
Given the similarity of the results, we conclude that our calibrated output cost function is quite in line with the data.

Finally, our model reproduces quantitatively standard empirical regularities of emerging economies: strong correlation between consumption and output, and volatility and countercyclicality of net exports. Along these dimensions, our model performs similarly to that of \cite{Chatterjee_2010} and the baseline model.

\subsection{A Graph for the Argentinean Case}
\label{sec:ARG}

\begin{figure}[htp]
  \centering
\includegraphics[height=3.0in,width=\textwidth]{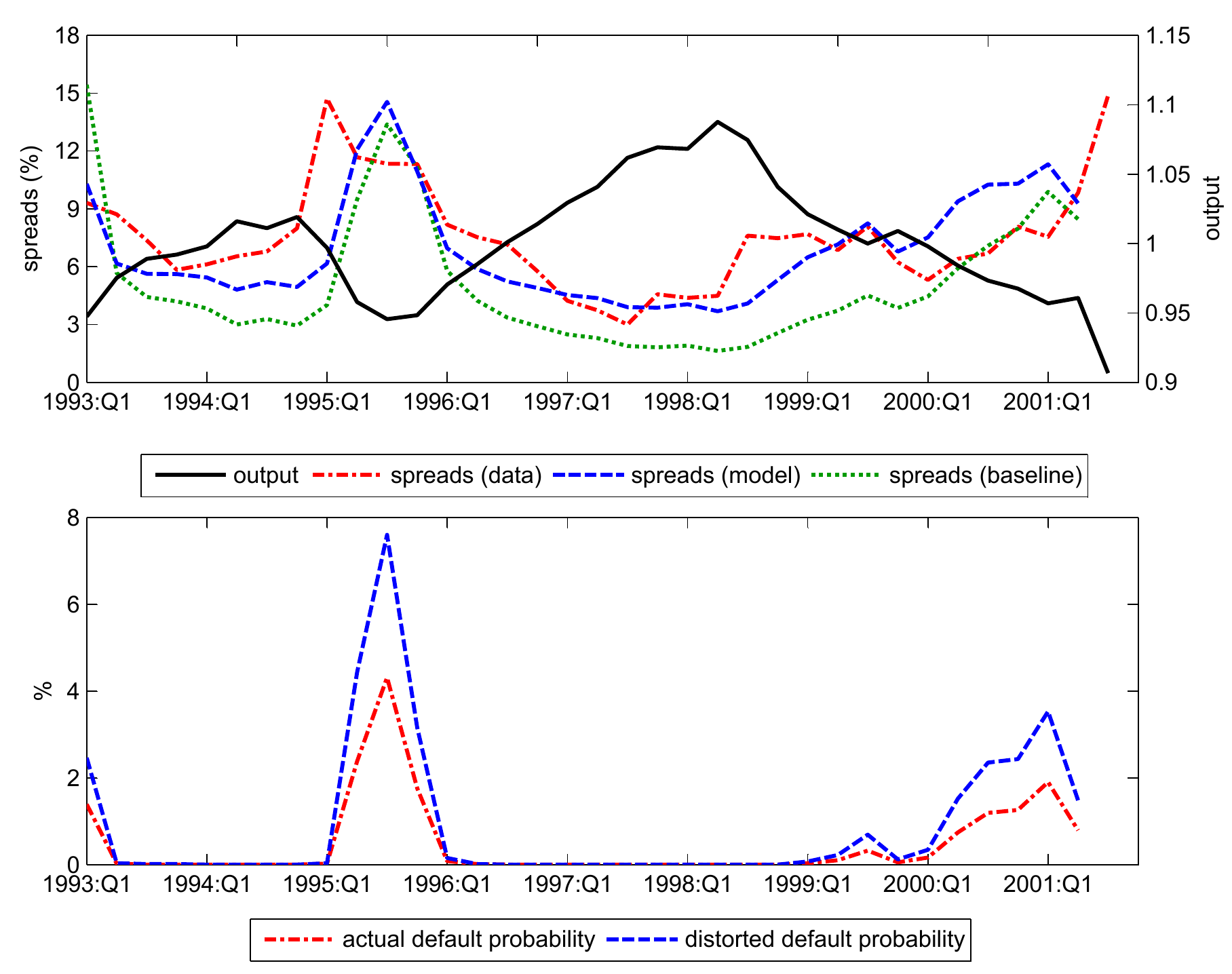}
  \caption{{\footnotesize{Top Panel: Output for Argentina; spreads generated by our long-term debt model and the baseline model; and actual spreads (measured by the EMBI+). Bottom Panel: One-step-ahead conditional probabilities of default under the distorted model and the approximating model.}}}
\label{fig:Fdata_ARG}
\end{figure}

In order to showcase the dynamics generated by long-term debt model, we perform the following exercise. We input into the model the output path observed in Argentina for 1993:Q1 to 2001:Q4. Given this and an initial level of debt, the model generates a time series for the annualized spread and for one-step-ahead conditional probabilities of default under both the approximating and distorted models. Figure \ref{fig:Fdata_ARG} depicts the results. 
The top panel shows the output path, jointly with the time series for bond spreads exhibited in the data and delivered by our model. For comparison, we also plot the spreads generated by each of the baseline models. The bottom panel displays the conditional default probabilities according to our model.


%

Our model does a better job matching the actual spreads than the corresponding baseline model. The difference between the spreads generated by the model can be largely explained by the behavior over time of one-step-ahead conditional probabilities of default. While we observe zero or negligible default risk right before and after the year of 1995, a salient feature of long-term debt models is that they can generate considerable bond spreads even in the absence of default risk in the near future. Lenders typically demand high returns on their long-term debt holdings to get compensated for possible capital losses due to future defaults on the unmatured fraction of their bonds. Additionally, further compensation is required for potential drops in the future market value of outstanding bonds as the borrower might dilute its debt.

In any case, while non-negligible, the subjective probability of default next period is higher than the actual one. More importantly, the wedge between the two probabilities is greater when output is low (and default is more likely next period),  e.g. see results for 1995:Q2 to 1995:Q4 and from 2000:Q2 onwards.

Finally, it is worth pointing out that our results are consistent with the findings by \cite{Zhang}. Using CDS price data on Argentinean sovereign debt at daily frequency from January 1999 to December 2001, \cite{Zhang} estimates a three-factor credit default swap model and computes the implied one-year physical and risk-neutral default probabilities. In line with our results, his risk-neutral default probability is always higher than its physical counterpart, and the wedge between them is time-varying, and typically increases with the physical default probability.


\section{Measuring Model Uncertainty}
\label{sec:dep}

In this section we present two different procedures to measure the amount of model uncertainty in this economy and interpret the
value of penalty parameter $\theta$ in the calibration. The first one is the Detection error probability (DEP) procedure used in \cite{AHS}, \cite{Maenhout}, and \cite{BHS} among others. The second one is, to our knowledge, a novel procedure which allows the researcher to focus on specific aspects of the probability distribution implied by the model.

\subsection{Detection Error Probabilities}



\begin{figure}[htp]
	\centering
	\includegraphics[height=3.0in,width=\textwidth]{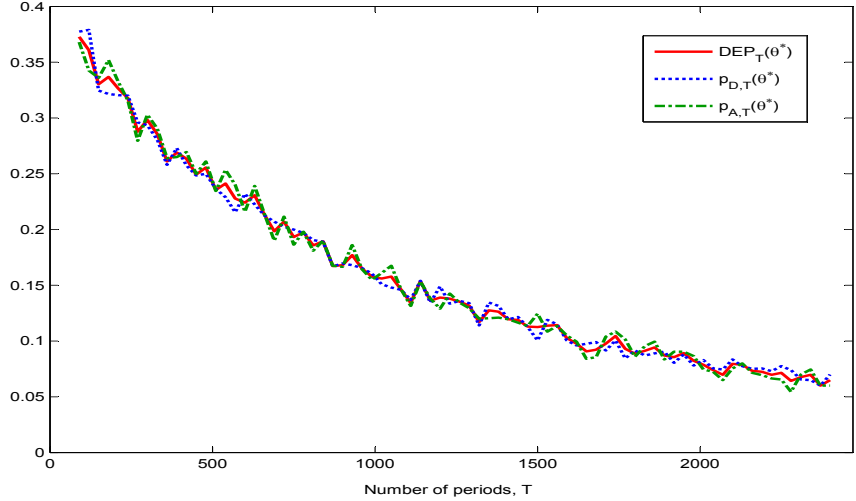}
	\caption{Detection error probability and its components as function of the number of periods, T, for our calibrated economy.}
	\label{fig:dep}
\end{figure}

Let $L_{A,T}$ and $L_{\theta,T}$ be the likelihood functions corresponding to the
approximating and distorted models for $(Y_{t})_{t=1}^{T}$, respectively. Let $Pr_{A}$ and $Pr_{\theta}$ be the respective probabilities over the data,
generated under the approximating and distorted models. Let $p_{A,T}(\theta) \equiv Pr_{A} \left(  \log \left(  \frac{L_{\theta,T}}{L_{A,T}}   \right) > 0  \right)$ be the probability the likelihood ratio test indicates that
the distorted model generated the data (when the data were generated by the approximating model). We define $p_{D,T}(\theta) \equiv Pr_{\theta} \left(  \log \left(\frac{L_{\theta,T}}{L_{A,T}} \right) < 0   \right)$ similarly. Finally, let the DEP be obtained by averaging $p_{D,T}(\theta)$ and $p_{A,T}(\theta)$:
\begin{equation*}
	DEP_{T}(\theta)=\frac{1}{2}\left( p_{A,T}(\theta)+p_{D,T}(\theta) \right).
\end{equation*}
If the two models are very similar to each other, mistakes are likely, yielding high values of $p_{A}(\theta)$ and $p_{D}(\theta)$; the opposite is true if the models are not similar.\footnote{The weight of one-half is arbitrary; see \cite{BHS} among others. Moreover, as the number of observations increases, the weight becomes less relevant, since the quantities $p_{A,T}$ and $p_{D,T}$ get closer to each other; as shown in Figure \ref{fig:dep}}

The aforementioned quantities can be approximated by means of simulation. We start by setting an initial debt level and endowment vector. We then simulate time series for output for $T'=2,000 + T$ periods (quarters), where $T=240$.\footnote{To make our results for the DEP comparable with those of \cite{BHS} and \cite{Bidder_Smith}, we consider a similar number of periods and thereby T=240 is chosen. If instead T was set to replicate the number of periods used in the calibration, the DEP would be considerably higher for the same probability distortions. For both models, we ignore the first 2,000 observations in order to avoid any dependence on our initial levels of debt and endowments.} The process is repeated 2,000 times. For each time-series realization, we construct $L_{A,T}$ and $L_{\theta,T}$. We then compute $p_{A,T}(\theta)$ as
the percentage of times the likelihood ratio test indicates that
the worst-case model generated the data (when the data were generated by the approximating model).\footnote{In the case of $L_{A} = L_{\theta}$ we count this as a false rejection with probability $0.5$.} We construct $p_{D,T}(\theta)$ analogously.

For a \emph{given} number of observations (in our case $74$), as
$\theta \rightarrow +\infty$, the approximating and distorted models
become harder to distinguish from each other and the detection error
probability converges to $0.5$. If instead they are distant from each other, the detection error
probability is below $0.5$, getting closer to $0$ as the discrepancy between
the models gets larger.

Following \cite{BHS} we consider a threshold for the DEP of $0.2$; values of $DEP_{T}(\theta)$ that are larger or equal are deemed acceptable. In our calibration, our DEP is above this threshold, since $DEP_{T}(\theta) = 0.313$. For this value of $\theta$ (and other parameters), $p_{A,T} = 0.306$ and $p_{D,T} = 0.321$. So the weight of $0.5$ does not play an important role.

We conclude the section by proposing an alternative view for
interpreting $\theta$. This is based on the following
observation: for any \emph{fix} finite $\theta$ (for which $L_{\theta,T}$ exists), $L_{A,T} \ne
L_{\theta,T}$ with positive probability; thus, as the number of
observations increases, $p_{T,k}(\theta) \rightarrow 0$ for $k=\{A,D\}$. Therefore, for a given level
of $\theta$ and an a priori chosen level $\alpha \in (0,1)$, which
does not depend on $\theta$, we can define $T_{\alpha,\theta} \equiv \max\{ T \colon p_{T}(\theta) = \alpha \}$, as the maximum
number of observations before $DEP_{T}(\theta)$ falls bellow
$\alpha$. A heuristic interpretation of this number is that the agents need at least $T_{\alpha,\theta}$ observations to be able to distinguish between the two models at a certainty level of $\alpha$. The higher this number, the harder it is to distinguish
between the models.

Figure \ref{fig:dep} plots $\{p_{T,A}(\theta^{\ast}),p_{T,D}(\theta^{\ast}),DEP_{T}(\theta^{\ast})\}_{T=90}^{2,400}$
for $\theta^{\ast} = 0.619$, the value of $\theta$ in our calibration. For a level of $\alpha = 0.2$, we see that $T_{\alpha,\theta^{\ast}}
\approx 700$. That is, one needs
approximately 9.5 times our sample of $74$, in order to obtain a level
of $\alpha=0.2$ for $DEP_{T}(\theta^{\ast})$ and consequently claim that these models are sufficiently different from each other, according to this criterion.

\subsection{Moment Based Uncertainty Measure}

The DEP criterion compares the \emph{likelihood} implied by each model. In our setting, however, as suggested by Figure \ref{fig:2_densities}, we expect the probabilities under the approximating and distorted models to differ mainly in the lower tail of the domain, where default predominantly occurs.  We thus propose a measure of model uncertainty which allow us to focus on such particular features of the probabilities models. We achieve this by constructing the measure using a GMM-based criterion function, which let us analyze these particular features of the probability distribution through a chosen vector of moments.\footnote{Extending the analogy to DEP, one can view DEP as a measure based on the likelihood ratio.}

Formally, we first define the following function: given any (stationary) distribution over $(Y_{t})_{t}$, $P$, let $\boldsymbol{\nu}(P) \in \mathbb{R}$ be the \emph{parameter of $P$}. That is, $\boldsymbol{\nu}(P)$ summarizes the features of the probability of the data we want to focus on. The parameter functions we consider are such that there exists a function $g : \mathbb{Y} \times \mathbb{R} \rightarrow \mathbb{R}$ such that $E_{P}[g(Y,\boldsymbol{\nu}(P))] = 0$. That is, the parameter is identified by a moment condition given by $g$.\footnote{It is straightforward to extend our setup to allow for vector-valued $g$.}

In our setup, due to the fact that default occurs predominantly for low values of the endowment, we are interested in the $\tau$-quantile of the distribution, i.e., $\boldsymbol{\nu}(P)$ such that $E_{P} [ 1\{  Y \leq  \boldsymbol{\nu}(P)  \}   ] = \tau $ and thus $g(y,\nu)= 1\{  Y \leq  \nu  \}  - \tau$. In particular, we choose $\tau = 0.1$ because around 70 percent of the default episodes in our model occur for endowment realizations below the associated level $\boldsymbol{\nu}(P)$; i.e., the set $\{ Y \leq  \boldsymbol{\nu}(P)  \}$ is a very good approximation of the set where most of the defaults occur.\footnote{In our case $\boldsymbol{\nu}$ is real-valued, but our analysis can easily be extended to the case where $\boldsymbol{\nu}$ is vector-valued.}

Given data $(Y_{t})_{t=1}^{T}$, let $Q_{T}(P) = (T^{-1} \sum_{t=1}^{T} g(Y_{t},\boldsymbol{\nu}(P))  )^{2} V $ where $V$ is a positive number. That is, $Q_{T}(P)$ is the (sample) GMM criterion function associated to the moments determined by $g$ and $\boldsymbol{\nu}$.

\begin{figure}[htp]
	\centering
	\includegraphics[height=3.0in,width=\textwidth]{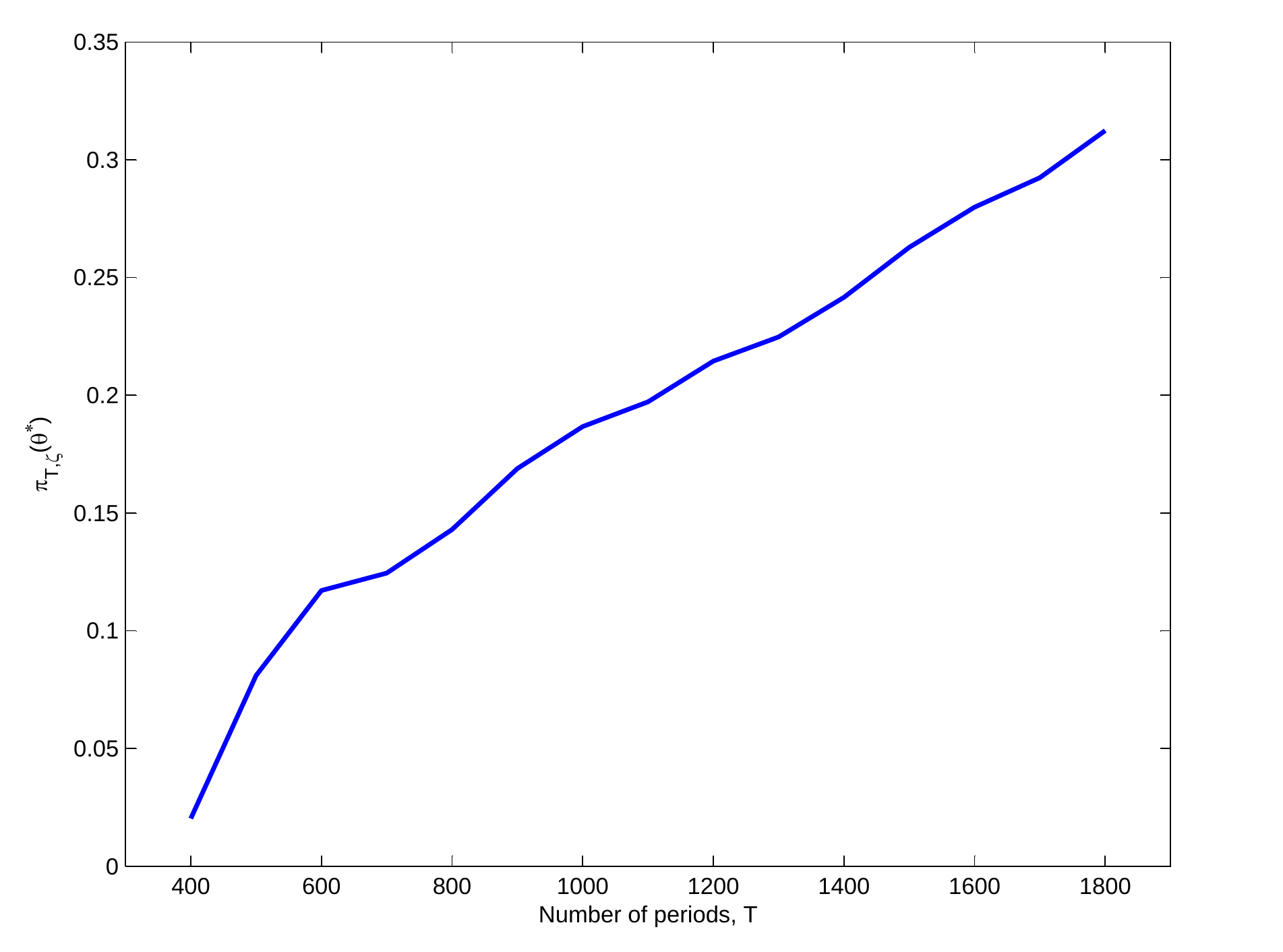}
	\caption{The uncertainty measure $\pi_{T,\zeta}(\theta^{\ast})$ as function of the number of periods, T, for our calibrated economy.}
	\label{fig:FpiT}
\end{figure}

The value of $V$ is chosen such that when data is drawn from $Pr_{A}$,\footnote{Under mild assumptions over $Pr_{A}$ which ensure the validity of the Central Limit Theorem, such $V$ always exists.}
\begin{align}
	T \times  Q_{T}(Pr_{A}) \Rightarrow \chi^{2}_{1}.
\end{align}
For any $\zeta \in (0,1)$, let $c_{\zeta}$ be the $(1-\zeta)$-quantile of $\chi^{2}_{1}$, and \begin{align*}
	\pi_{T,\zeta}(\theta) = Pr_{A} \left(  T \times Q_{T}(Pr_{\theta})  \geq c_{\zeta}  \right).
\end{align*}
Clearly, as $T \rightarrow \infty$, $\pi_{T,\zeta}(\theta) \rightarrow 1$ for any finite $\theta$. However, for our choice of $g$, if $\theta$ is such that $Pr_{\theta}(Y \leq \boldsymbol{\nu}(Pr_\theta)) \approx Pr_{A}(Y \leq \boldsymbol{\nu}(Pr_\theta))$ then, even for moderately high values of $T$, we should expect $\pi_{T,\zeta}(\theta) \approx \zeta $.

In light of this remark, and similarly to DEP, given a $T$ and a $\zeta \in (0,1)$, the researcher needs to stipulate a threshold larger or equal than $\zeta$ for which values of $\pi_{T,\zeta}(\theta)$ \emph{below} the threshold are deemed acceptable. That is, if $\theta$ is such that $\pi_{T,\zeta}(\theta)$ is below the threshold, then $Pr_{\theta}$ is considered to be ``close'' to $Pr_{A}$ and hence cannot be distinguished from each other given the observations available. Alternatively --- also in line with what we proposed for DEP ---, the researcher can choose a value of $\alpha \in [\zeta,1)$ and construct $T_{\alpha,\theta} = \max \{ T :  \pi_{T,\zeta}(\theta) \leq \alpha  \}$.

For our choice of $\boldsymbol{\nu}(Pr_{\theta})$, given by the 0.1-quantile of the distribution $Pr_{\theta}$, we construct $\pi_{T,\zeta}$ for different values of $T$.\footnote{To do so, we first approximate $\boldsymbol{\nu}(Pr_{\theta})$ using 100,000 draws from $Pr_{\theta}$. Then, for different values of $T$, we run 50,000 simulations of $T$ observations each under the approximating model $Pr_{A}$ to compute $\pi_{T,\zeta}$. In all cases, additional 1,000 observations were drawn at the beginning of each simulation and were disregarded to avoid dependence on initial values. }

The results we obtain are for $\zeta = 0.05$. For a threshold of $0.05$ and for $T=240$ (the number of periods used in the computation of the DEP), our calibrated value of $\theta$, $\theta^{\ast} = 0.619$, yields an admissible value of $\pi_{T,0.05}(\theta^{\ast})=0.028$. 
Figure \ref{fig:FpiT} plots $\{\pi_{T,0.05}(\theta^{\ast})\}_{T=400}^{2,400}$.\footnote{The fact that $T$ starts at $400$ is not crucial, but considering values of $T$ that are too small, may yield unreliable results due to poor small sample behavior or $\pi_{T,\zeta}$.} We can see that for thresholds, $\alpha$, of $0.05$ or $0.10$, the agent needs at least (approx.) $430$ or $580$ observations to distinguish the models with the desired certainty.

These values are smaller than those obtained for DEP, reflecting the fact that our measure focuses on the low values of the endowment where the approximated and distorted models differ the most; i.e. in this sense our measure is more stringent than DEP. They are, however, still larger than our sample of $74$ quarters. Thus, even also under this new measure of uncertainty we view our value of $\theta$ as entailing a quite conservative amount of model uncertainty.


\section{Conclusion}
\label{sec:conc}

This paper addresses a well-known puzzle in the sovereign default literature: why
are bond spreads for emerging economies so high if default episodes are rare events, with a low probability of occurrence?
Using \cite{Eaton} general equilibrium framework, extended by \cite{Chatterjee_2010} to allow for long-term debt, we provide an explanation to resolve this puzzle based on concerns about model misspecification.

In recent years, some emerging economies such as Argentina, South Africa, Brazil, Colombia, and Turkey have issued GDP-indexed or inflation-indexed sovereign bonds. Credibility of the sovereign and transparency of its statistic agency are paramount for the success of these markets. Our framework suggests that potential concerns of misreporting output growth or inflation would be priced in by investors when lending to the sovereign, and might question the desirability of these policies.

\bibliographystyle{aea}
\bibliography{mybib_ROBDEF}

\newpage

\appendix



\section{Irrelevance of Endowment for Investors}
\label{sec:yL}

In what follows we describe in more detail the environment presented in Lemma \ref{lem:lenders_size}. In fact, we allow for a more general specification for the process of the lenders' endowment is assumed and we allow the lender to distrust it.

We assume a more general stochastic process for the endowment of the lender $z_{t} \in \mathbb{Z}$, namely
\begin{align}\label{eqn:len-endow}
Z_{t+1} = \rho_{0} + \rho_{1} Z_{t} + \epsilon_{t+1},
\end{align}
where $\epsilon_{t+1}$ is distributed according to the cdf $F_{\epsilon}(\cdot | y_{t+1},y_{t})$. Note that under this specification the endowments of the borrower and of the lenders can be correlated. For the ease of notation, we assume that both $Y_{t}$ and $Z_{t}$ are continuous random variables with conditional pdfs $f_{Y}$ and $f_{Z}$, respectively. Also, for simplicity we omit $X_{t}$; the generalization that allows for it is straightforward.

We also allow the lenders to distrust the specification of the stochastic process of  $Z$, as well as that of $Y$, but possibly to a different extent. Let $\theta$ and $\eta$ be the penalty parameters controlling for the degrees of concern about model misspecification for the distributions of $y$ and $z$, respectively. Different degrees of concern for each process are consistent with our view that there are more extensive, reliable datasets, especially from official statistical sources, containing relevant macro-financial information for developed economies and global capital markets, than for emerging economies.



In this economy, to distort the expectation of the lenders' continuation values, the minimizing agent will be placing two types of probability distortions, albeit not simultaneously. Indeed, first, it distorts the distribution of $Z_{t+1}$ for each realization of $Y_{t+1}$. Then, taking the resulting distorted continuation values for the lender as given, the minimizing agent proceeds to twist the probability of $Y_{t+1}$.

For convenience, we define the following risk-sensitive operators: $R_{\theta}$ and $T_{\eta}$, where for any $g \in L^{\infty}(\mathbb{Y})$,
\begin{align}
\label{eq:7}
R_{\theta}[g](y) = - \theta \log E_{Y} \left[ \exp \left\{- \frac{g(Y')}{\theta}          \right\}  |y \right]
\end{align}
for any $y$; and for any $h \in L^{\infty}(\mathbb{Z})$,
\begin{align}
\label{eq:8}
T_{\eta}[h](y',y,z) = - \eta \log E_{Z} \left[ \exp \left\{-\frac{h(y',Z^{\prime})}{\eta}          \right\}  |y',y,z \right]
\end{align}
for any $(y',y)$, and where $E_{Z}[\cdot |y',y,z]$ is the conditional expectation of $Z_{t+1}$, given $(y_{t+1},y_{t},z_{t})=(y',y,z)$.


\begin{theorem}\label{thm:pricing}
	There exists a recursive equilibrium for this economy such that the equilibrium price function is given by:
	\begin{align}
	\label{eq:1}
	q^{o}(y_{t},B_{t+1}) =  \gamma E_{Y} \left[ (\lambda + (1-\lambda)(\psi+q^{o}(Y_{t+1},B^{o}(W_{t+1},B_{t+1})))) \delta^{o}(Y_{t+1},B_{t+1})  \mathbf{m}^{o}(Y_{t+1};y_{t},B_{t+1})  \right]
	\end{align}
	for any $(y_{t},B_{t+1})$, where: (i) For any $y_{t+1}$,
	\begin{align}
	\label{eq:2}
	\mathbf{m}^{o}(y_{t+1};y_{t},B_{t+1}) \equiv \frac{\exp \left\{ - \frac{ \frac{T_{\eta(1-\gamma \rho_{1})}[\epsilon_{t+1} ](y_{t+1},y_{t})}{1-\gamma \rho_{1}} + \bar{W}^{o}(y_{t+1},B_{t+1},B_{t+1}) }{\theta}\right\} }{E_{Y} \left[ \exp \left\{ - \frac{ \frac{T_{\eta(1-\gamma \rho_{1})}[\epsilon_{t+1} ](Y_{t+1},y_{t})}{1-\gamma \rho_{1}} + \bar{W}^{o}(Y_{t+1},B_{t+1},B_{t+1}) }{\theta} \right\} |y_{t} \right] }
	\end{align}
	(ii) $(B^{o},\delta^{o})$ correspond to the optimal policy functions in the borrower's problem, given $q^{o}$; and (iii)
	\begin{align}
	\bar{W}^{o}(y,B,B) \equiv \delta^{o}(y,B)\bar{W}^{o}_{R}(y,B,B) +  (1-\delta^{o}(y,B))\bar{W}^{o}_{A}(y),
	\end{align}
	where $(\bar{W}^{o}_{R},\bar{W}^{o}_{A})$ solve the following problem
	\begin{align}
	\notag
	\bar{W}^{o}_{R}(y_{t},B_{t+1},b_{t}) = & \max_{b_{t+1}}\left\{ \{ q^{o}(y_{t},B_{t+1})(b_{t+1} - (1-\lambda)b_{t} - (\lambda + (1-\lambda)\psi)b_{t}\} \right.\\ \label{eq:3}
	& \left.  + \gamma R_{\theta} \left[   \frac{ T_{\eta(1-\gamma \rho_{1})}[\epsilon_{t+1} ](Y_{t+1},y_{t})}{1-\gamma \rho_{1}} + \bar{W}(Y_{t+1},B_{t+1},b_{t+1})  \right] (y_{t}) \right\},
	\end{align}
	and
	\begin{align}
	\label{eq:4}
	\bar{W}^{o}_{A}(y_{t}) = \gamma R_{\theta} \left[   \frac{ T_{\eta(1-\gamma \rho_{1})}[\epsilon_{t+1} ](Y_{t+1},y_{t})}{1-\gamma \rho_{1}} + (1-\pi)\bar{W}_{R}(Y_{t+1}) + \pi \bar{W}(Y_{t+1},0,0)  \right] (y_{t}),
	\end{align}
\end{theorem}

We relegate the somewhat long proof to the end of this section. A few remarks about the theorem are in order. First, the borrower's optimal policy functions $(\bar{W}^{o}_{R},\bar{W}^{o}_{A})$ do not depend on $z_{t}$. This is because the price function does not depend on $z_{t}$ and thus the borrower does not need to keep track of it in order to predict future prices.

Second, by inspection of equation (\ref{eq:1}) we can formulate the following corollary

\begin{corollary}
	If $\epsilon_{t+1}$ is independent of $(Y_{t})_{t}$, i.e., $F_{\epsilon}(\cdot | y_{t+1},y_{t}) = F_{\epsilon}(\cdot)$, then $q^{o} = q$ and $(B^{o},\delta^{o}) = (B,\delta)$.
\end{corollary}

That is, if $\epsilon_{t+1}$ is independent of $(Y_{t})_{t}$, then the equilibrium price function and debt and default decisions are \emph{identical} to those in our economy; this corollary clearly includes the lemma in the text as a particular case.

\subsection{Proof} 
\label{sec:lemma_gen}

Analogously to Section \ref{sec:LENDERS_PREF}, lenders' utility over consumption plans $\mathbf{c}^{L}$ after any history $(t,y^{t},z^{t})$ is henceforth given by
\begin{align}\label{eqn:BE_MA_0}
U_{t}(\mathbf{c}^{L};y^{t},z^{t}) =  c^{L}_{t}(y^{t},z^{t}) + \gamma \min_{(m,n) \in \mathcal{M} \times \mathcal{N}}  \left\{ \theta \mathcal{E}_{\theta}[m](y_{t}) +  E_{Y} \left[m(Y_{t+1}) \left[\eta \mathcal{E}_{\eta}[n](z_{t}) \right. \right. \right. \\
\qquad   \left. \left.+  n(Z_{t+1})U_{t+1}(\mathbf{c}^{L};y^{t},Y_{t+1},z^{t},Z_{t+1}) \mid y_{t},z_{t} \right]  \right\},   \notag
\end{align}
where the conditional relative entropies $\mathcal{E}_{\theta} : \mathcal{M} \rightarrow \{ g : \mathbb{Y} \rightarrow \mathbb{R}_{+} \}$ and $\mathcal{E}_{\eta} : \mathcal{N} \rightarrow \{ g : \mathbb{Z} \rightarrow \mathbb{R}_{+} \}$ are defined in analogy to (\ref{eqn:entropy}).

By similar calculations to those in Section \ref{app:PO_MA} of the Supplementary Material, one can show that the corresponding Bellman equation is
\begin{align*}
W_{R}(v_{t},B_{t},b_{t}) =  \min_{(m,n) \in \mathcal{M} \times \mathcal{N}} \max_{b_{t+1}}&\left\{ z_{t} + G(b_{t},b_{t+1};B_{t+1},v_{t})  \right. + \theta \gamma \mathcal{E}[m](y_{t}) +  \eta \gamma E_{Y}[m(Y_{t+1})\mathcal{E}[n](z_{t})|y_{t}] \\
&\qquad+ \left.\gamma  E_{V} \left[ m(Y_{t+1}) n(Z_{t+1}) W(V_{t+1},B_{t+1},b_{t+1}) |v_{t} \right] \right\},
\end{align*}
where let $(z_{t},b_{t},b_{t+1};B_{t+1},v_{t}) \mapsto z_{t} + G(b_{t},b_{t+1};B_{t+1},v_{t})  \equiv z_{t}+q(v_{t},B_{t+1})(b_{t+1}-(1- \lambda) b_{t}) -(\lambda + (1-\lambda)\psi)b_{t}$ be the per-period payoff. The expression for $W_{A}$ is analogous.

	Let $v_{t} \equiv (z_{t},y_{t})$. The proof consists of two parts. First, assuming that $q(v_{t},B_{t+1}) = q(y_{t},B_{t+1})$ we show that $W_{i}(v_{t},B_{t},b_{t}) = A_{0} + A_{1} z_{t} + \bar{W}_{i}(y_{t},B_{t},b_{t})$ for all $i \in \{ R,A\}$ where
	\begin{align*}
	A_{0} = \frac{\rho_{0} A_{1} }{1-\gamma } ~and~A_{1} = \frac{1}{1-\gamma \rho_{1}}.
	\end{align*}
	Then, given this result, we prove that the equilibrium price function from the FONC of the lender's problem is in fact $q(v_{t},B_{t+1}) = q(y_{t},B_{t+1})$. This shows that the equilibrium mapping that maps prices into prices, in fact maps functions of $(y_{t},B_{t+1})$ onto themselves, and thus the equilibrium price must have this property.
	
	Observe that, given $q(v_{t},B_{t+1}) = q(y_{t},B_{t+1})$, the borrower does not consider $z_{t}$ as part of the state, and thus $\delta(v_{t+1},B_{t+1}) = \delta(y_{t+1},B_{t+1})$. Hence,
	\begin{align*}
	W(v_{t+1},B_{t+1},b_{t+1}) = & \delta (y_{t+1},B_{t+1})W_{R}(v_{t+1},B_{t+1},b_{t+1})+(1-\delta(y_{t+1},B_{t+1}))W_{A}(v_{t+1}) \\
	\equiv & \bar{W}(y_{t+1},B_{t+1},b_{t+1})  + A_{0} + A_{1} z_{t+1} .
	\end{align*}
	
	Given the assumption on prices, it follows that
	{\small{\begin{align*}
			W_{R}(v_{t},B_{t},b_{t}) = &\min_{(m,n) \in \mathcal{M} \times \mathcal{N}} \max_{b_{t+1}}\left\{ z_{t} + G(b_{t},b_{t+1};B_{t+1},y_{t}) + \theta \gamma \mathcal{E}[m](y_{t}) \right. \\
			& \left.  +  \eta \gamma E_{Y}[m(Y_{t+1})\mathcal{E}[n](z_{t})|y_{t}] + \gamma  E_{V} \left[ m(Y_{t+1}) n(Z_{t+1}) W(V_{t+1},B_{t+1},b_{t+1}) |v_{t} \right] \right\},
			\end{align*}}}
	where $\mathcal{N}$ is defined similarly to $\mathcal{M}$.
	
	Solving to the minimization problem yields
	\begin{align*}
	&  W_{R}(v_{t},B_{t},b_{t})\\
	& = \max_{b_{t+1}}\left\{ z_{t} + G(b_{t},b_{t+1};B_{t+1},y_{t})  -  \theta \gamma  \log E_{Y} \left[ \exp \left\{ - \frac{ T_{\eta}[W(\cdot,Y_{t+1},B_{t+1},b_{t+1})](Y_{t+1},z_{t}) }{\theta}\right\} | y_{t} \right]   \right\}.
	\end{align*}
	
	By assumption over $W$, we have that
	{\footnotesize{\begin{align*}
			&  W_{R}(v_{t},B_{t},b_{t}) = \bar{W}_{R}(y_{t},B_{t},b_{t}) +  A_{1} (z_{t}) +  A_{0} \\
			= & \max_{b_{t+1}}\left\{   G(b_{t},b_{t+1};B_{t+1},y_{t})  -  \theta \gamma  \log E_{Y} \left[   \exp \left\{ - \frac{ A_{1} T_{\eta/A_{1}}[\epsilon_{t+1} ](Y_{t+1},y_{t}) + \bar{W}(Y_{t+1},B_{t+1},b_{t+1}) }{\theta}\right\} |y_{t} \right] \right\} \\
			&  + (\gamma A_{1} \rho_{1} +1 )(z_{t}) + \gamma (A_{0}  + A_{1} \rho_{0}).
			\end{align*}}}
	Therefore, it must be the case that $A_{1} = (\gamma A_{1} \rho_{1} +1 ) ~and~ A_{0} = \gamma (A_{0}  + A_{1} \rho_{0})$. Similar algebra for $W_{A}$ yields
	{\footnotesize{\begin{align*}
			\bar{W}_{A}(y_{t}) +& A_{0} + A_{1} z_{t} =  \gamma \left\{  - \theta \log E_{Y} \left[   \exp \left\{ - \frac{\{(1-\pi
				)\bar{W}_{A}(Y_{t+1})+\pi \bar{W}(Y_{t+1},0,0)\} + A_{1}  T_{\eta/A_{1}}[\epsilon_{t+1}](Y_{t+1})}{\theta} | y_{t} \right] \right\}  \right\}\\
			& \qquad +  \gamma (A_{0} + A_{1} \rho_{0}) + (1+\gamma A_{1} \rho_{1}) z_{t}.
			\end{align*}
		}}
		Therefore, the same solution for $A_{0}$ and $A_{1}$ holds for $W_{A}$. Hence,
		\begin{align*}
		\bar{W}_{R}(y_{t},B_{t},b_{t}) = \max_{b_{t+1}}   &G(b_{t},b_{t+1};B_{t+1},y_{t})  \\
		&-\theta \gamma  \log E_{Y} \left[ \exp \left\{ - \frac{ \frac{T_{\eta(1-\gamma \rho_{1})}[\epsilon_{t+1} ](Y_{t+1},y_{t})}{1-\gamma \rho_{1}} + \bar{W}(Y_{t+1},B_{t+1},b_{t+1}) }{\theta}\right\} | y_{t} \right].
		\end{align*}
		
		The FONC and envelope conditions for $b_{t+1}$ (assuming interior solution) yield
		{\footnotesize{\begin{align*}
				q^{o}(y_{t},B_{t+1}) = \gamma E_{Y} \left[ \Upsilon(W_{t+1},B_{t+1}) \delta^{o}(Y_{t+1},B_{t+1}) \frac{\exp \left\{ - \frac{ \frac{T_{\eta(1-\gamma\rho_{1})}[\epsilon_{t+1} ](Y_{t+1},y_{t})}{1-\gamma \rho_{1}} + \bar{W}(Y_{t+1},B_{t+1},B_{t+1}) }{\theta}\right\} }{E_{Y} \left[ \exp \left\{ - \frac{ \frac{T_{\eta(1-\gamma \rho_{1})}[\epsilon_{t+1} ](Y_{t+1},y_{t})}{1-\gamma \rho_{1}} + \bar{W}(Y_{t+1},B_{t+1},B_{t+1}) }{\theta}\right\}  \Big| y_{t}  \right] } \Bigg| y_{t}   \right],
				\end{align*}}}
		where $\Upsilon(w',B') \equiv \Big(\lambda + (1-\lambda)\big(\psi+q^{o}(y',B^{o}(w',B'))\big)\Big) $.

\newpage

\clearpage
\setcounter{page}{1}
\setcounter{section}{0}
\renewcommand\thesection{S.\arabic{section}}

\section{Online Supplementary Material}

\section{Recursive Formulation of the Problem of the ``Minimizing Agent''}
\label{app:PO_MA}

In this section, we show that the principle of optimality holds for the Problem of the ``minimizing agent''.

Let $\mathbf{c}^{L}$ be a consumption plan. A feasible consumption plan is one that satisfies the budget constraint for each $t$. Preferences over consumption plans for lenders are then
described as follows. For any given consumption plan $\mathbf{c}^{L}$ and initial state $w_{0}$,
the lifetime utility over such plan is given by\footnote{Without the i.i.d. component $x_{t}$, the lifetime utility for the lender over $\mathbf{c}^{L}$ would simply be given by
	\begin{align}
	U_{0}(\mathbf{c}^{L};y_{0}) \equiv \min_{(m_{t+1})_{t}} \sum_{t=0}^{\infty}
	\gamma^{t} E \left[  M_{t}(Y^{t})\{  c^{L}_{t}(Y^{t})   + \theta \gamma
	\mathcal{E}[m_{t+1}(\cdot|Y^t)](Y_{t}) \} \mid y_{0} \right].
	\end{align}
}
\begin{align}\label{eqn:SP_MA}
&U_{0}(\mathbf{c}^{L};w_{0}) \equiv \min_{(m_{t+1})_{t}} \sum_{t=0}^{\infty}
\gamma^{t} E \left[  M_{t}(W^{t})\left( c^{L}_{t}(W^{t}) + \theta \gamma
\mathcal{E}[m_{t+1}(\cdot|W^{t})](Y_{t}) \right) \mid w_{0} \right]&\\
& \qquad E_{Y}[m_{t+1}(Y_{t+1}|W^{t})\mid y_{t}] =1, \notag&
\end{align}
where $E$ denotes the expectation with respect to $W^{t}$ under the probability measure $P$, $\gamma \in (0,1)$ is the discount factor, the parameter $\theta \in (\underline{\theta },+\infty ]$ is a penalty parameter
that measures the degree of concern about model misspecification, and the mapping $\mathcal{E} : \mathcal{M} \rightarrow L^{\infty}(\mathbb{Y})$, with $\mathcal{M}$ defined in Subsection \ref{sec:LENDERS_PREF}, is the \emph{conditional relative entropy}, given by (\ref{eqn:entropy}).

We note that, since $\mathbb{B}$ is bounded, and, in equilibrium, $q_{t} \in [0,\gamma]$; any feasible consumption plan is bounded, i.e., $|c^{L}_{t}(W^{t})| \leq C < \infty$ a.s.

\begin{definition}
	Given a feasible consumption plan $\mathbf{c}^{L}$, for each $(t,w^{t})$, we say functions $(t,w^{t},\mathbf{c}^{L}) \mapsto \mathbb{U}_{t}(\mathbf{c}^{L};w^{t})$, satisfy the sequential problem of the ``minimizing agent'' (SP-MA) iff\footnote{Note that, since $c^{L}_{t} \geq -K_{0}$ and $\theta \mathcal{E} \geq 0$, the RHS of the equation is always well defined in $[-K_{0},\infty]$ where $K_{0}$ is some finite constant.}
	\begin{align}\notag
	&  \mathbb{U}_{t}(\mathbf{c}^{L};w^{t}) = \min_{(m_{t+j+1})_{j}} \sum_{j=0}^{\infty}
	\gamma^{j} E \left[  \left( \frac{M_{t+j}(W^{t+j})}{M_{t}(w^{t})} \right) \{  c^{L}_{t+j}(W^{t+j})  + \theta \gamma
	\mathcal{E}[m_{t+j+1}(\cdot|W^{t+j})](Y_{t+j}) \} |w_{t} \right],& \\
	& \qquad E_{Y}[m_{t+1}(Y_{t+1}|W^{t})\mid y_{t}] =1, & \label{eqn:SP_MA-1}
	\end{align}
	where $M_{t} \equiv \prod_{\tau=1}^{t} m_{\tau}$, $M_{0}=1$ and $E\left[\cdot | w^{t} \right]$ is the conditional expectation over all histories $W^{\infty}$, given that $W^{t} = w^{t}$.
\end{definition}

\begin{definition}
	Given a feasible consumption plan $\mathbf{c}^{L}$, for each $(t,y^{t})$, we say functions $(t,w^{t},\mathbf{c}^{L}) \mapsto U_{t}(\mathbf{c}^{L};w^{t})$, satisfy the functional problem of the ``minimizing agent'' (FP-MA) iff
	\begin{align}\label{eqn:BE_MA_3_app}
	U_{t}(\mathbf{c}^{L};w^{t}) =  c^{L}_{t}(w^{t}) + \gamma \min_{m_{t+1}(\cdot | w^{t}) \in \mathcal{M}} \left\{ E_{Y} \left[ m_{t+1}(Y_{t+1}|w^{t}) \mathcal{U}_{t+1}(\mathbf{c}^{L};w^{t},Y_{t+1}) \mid y_{t} \right] +\theta
	\mathcal{E}[m_{t+1}(\cdot|w^{t})](y_{t}) \right\},
	\end{align}
	where $\mathcal{U}_{t+1}(\mathbf{c}^{L};w^{t},y_{t+1}) = E_{X} [U_{t+1}(\mathbf{c}^{L};w^{t},X_{t+1},y_{t+1})]$.
\end{definition}

Henceforth, we assume that in both definitions, the ``min'' is in fact achieved. If not, the definition and proofs can be modified by using ``inf'' at a cost of making them more cumbersome. We also assume that $\mathbb{U}_{t}(\mathbf{c}^{L};\cdot)$, defined by (\ref{eqn:SP_MA-1}), is measurable with respect to $\mathcal{W}^{\infty}$.

\begin{theorem}\label{thm:PO_MA}
	For any feasible consumption plan $\mathbf{c}^{L}$,
	
	(a) If $({\mathbb{U}}_{t}(\mathbf{c}^{L};w^{t}))_{t,w^{t}}$ satisfies the SP-MA, then it satisfies the FP-MA.
	
	(b) Suppose there exist a function $(t,w^{t}) \mapsto \bar{U}_{t}(\mathbf{c}^{L};w^{t})$ that satisfy the FP-MA and
	\begin{align}\label{eqn:TC_MA}
	\lim_{T \rightarrow \infty} \gamma^{T+1} E \left[  M_{T+1}(W^{T+1}) \left( \bar{U}_{T+1}(\mathbf{c}^{L};w^{T+1})  \right) \mid w_{0} \right] = 0,
	\end{align}
	for all $M_{T+1}$ such that $M_{T+1} = m_{T+1} M_{T}$, $M_{0}=1$ and $m_{t+1} \in \mathcal{M}$. Then $(\bar{U}_{t}(\mathbf{c}^{L};w^{t}))_{t,w^{t}}$ satisfy the SP-MA.
\end{theorem}

The importance of this theorem is that it suffices to study the functional equation (\ref{eqn:BE_MA_3_app}). The proof of the theorem requires the following lemma (the proof is relegated to the end of the section).

\begin{lemma}\label{lem:min-M}
	In the program \ref{eqn:SP_MA-1}, it suffices to perform the minimization over $(m_{t})_{t} \in \mathbb{M}$ where
	\begin{align*}
	\mathbb{M} \equiv \left\{(m_{t})_{t} \colon m_{t} \in \mathcal{M} \cap \sum_{j=0}^{\infty}
	\gamma^{j}   E \left[ \left( \frac{M_{t+j}(W^{t+j})}{M_{t}(w^{t})} \right)
	\mathcal{E}[m_{t+j+1}(\cdot|W^{t+j})](Y_{t+j})  \} | w^{t} \right] \leq C_{C,\gamma,\theta}, \forall y^{t} \right\},
	\end{align*}
	where $C_{C,\gamma,\theta} = 2\frac{C}{(1-\gamma)\theta \gamma }$.
\end{lemma}

\bigskip

\begin{proof} \textsc{Proof of Theorem \ref{thm:PO_MA}.} Throughout the proof we use $E_{Y|X}$ to denote the expectation of random variable $Y$, given $X$.
	
	(a) From the definition of SP-MA and equation (\ref{eqn:SP_MA}), it follows that
	{\small{\begin{align*}
			&  \mathbb{U}_0(\mathbf{c}^{L};w_{0}) \\
			= & \min_{(m_{t+1})_{t}} \left\{ \{c^{L}_{0}(w_{0}) + \theta \gamma
			\mathcal{E}[m_{1}(\cdot|w_{0})](y_{0}) \} \right. \\
			& \left.+ \gamma  \sum_{t=1}^{\infty}
			\gamma^{t-1} E_{W^{1}|W_{0}} \left[ M_{1}(W^{1}) \left( E_{W^{t}|W^{1}} \left[ \frac{M_{t}(W^{t})}{M_{1}(W^{1})} \{  c^{L}_{t}(W^{t})  + \theta \gamma
			\mathcal{E}[m_{t+1}(\cdot|W^{t})](Y_{t}) \} \mid W_{1} \right] \right) | w_{0} \right]  \right\} \\
			= & \min_{(m_{t+1})_{t}} \left\{ \{ c^{L}_{0}(w_{0}) + \theta \gamma
			\mathcal{E}[m_{1}(\cdot|w_{0})](y_{0}) \} \right. \\
			(\star)  &+ \gamma E_{W^{1}|W_{0}} \left[ m_{1}(W^{1}) \left( \sum_{s=0}^{\infty}  \gamma^{s} E_{W^{s+1}|W^{1}} \left[ \frac{M_{s+1}(W^{s+1})}{M_{1}(W^{1})} \left\{  c^{L}_{s+1}(W^{s+1}) \right.\right.\right.\right.\\
			& + \left.\left.\left.\left.\left. \theta \gamma \mathcal{E}[m_{s+2}(\cdot|W^{s+1})](Y_{s+1}) \right\} | W^{1} \right] \right) \mid w_{0} \right]  \right\} \\
			&\geq  \min_{m_{1}} \left\{  c^{L}_{0}(w_{0}) + \theta \gamma
			\mathcal{E}[m_{1}(\cdot|w_{0})](y_{0}) \} + \gamma  E_{W^{1}|W_{0}}  [ m_{1}(Y^{1}|w_{0}) \left( \mathbb{U}_{1}(\mathbf{c}^{L};w_{0},W_{1})  \right) \mid w_{0} ]  \right\} \\
			&=  \min_{m_{1}} \left\{  c^{L}_{0}(w_{0}) + \theta \gamma
			\mathcal{E}[m_{1}(\cdot|w_{0})](y_{0}) \} + \gamma  E_{Y^{1}|Y_{0}} [ m_{1}(Y_{1}|w_{0}) E_{X}[\left( \mathbb{U}_{1}(\mathbf{c}^{L};w_{0},X_{1},Y_{1})  \right)] \mid w_{0} ]  \right\}.
			\end{align*}}}
	where  the first inequality follows from definition of $\mathbb{U}$. The step $(\star)$ follows from interchanging the summation and integral (we show this fact towards the end of the current proof).
	
	The final expression actually holds for any state $(t,y^{t})$,
	\begin{align}
	\mathbb{U}_{t}(\mathbf{c}^{L};w^{t}) & \geq  \min_{m_{t+1}} \left\{ c^{L}_{t}(w^{t}) + \theta \gamma
	\mathcal{E}[m_{t+1}(\cdot|w^{t})](y_{t}) \right. \label{eqn:BE_MA_1}\\
	&\qquad\left.+ \gamma E_{Y_{t+1}|Y_{t}}\left[ m_{t+1}(Y^{t+1}|w^{t}) E_{X}[\left( \mathbb{U}_{t+1}(\mathbf{c}^{L};w_{t},X_{t+1},Y_{t+1})  \right)]  \mid y_{t} \right]  \right\}. \notag
	\end{align}
	
	On the other hand, by definition of $\mathbb{U}$,
	{\footnotesize{\begin{align*}
			\mathbb{U}_0(\mathbf{c}^{L};w_{0})& \leq  M_{0}(w_{0}) \{c^{L}_{0}(w_{0}) + \theta \gamma
			\mathcal{E}[m_{1}(\cdot|w_{0})](y_{0}) \} \\
			& + \gamma  \sum_{t=1}^{\infty}  \gamma^{t-1}  E_{W^{1}|W_{0}} \left[ M_{1}(W^{1}) \left( E_{W^{t}|W^{1}} \left[ \frac{M_{t}(W^{t})}{M_{1}(W^{1})} \{  c^{L}_{t}(W^{t})  + \theta \gamma
			\mathcal{E}[m_{t+1}(\cdot|W^{t})](Y_{t}) \} | W^{1} \right] \right) \Big| w_{0} \right],
			\end{align*}}}
	for any $(M_{t})_{t}$ that satisfies the restrictions imposed in the text. In particular, it holds for $(M_{t})_{t}$ where $m_{1}$ is left arbitrary and $(m_{t})_{t\geq 2}$ is chosen as the optimal one. By following analogous steps to those before, it follows that
	\begin{align*}
	\mathbb{U}_0(\mathbf{c}^{L};w_{0}) \leq \{c^{L}_{0}(w_{0}) + \theta \gamma
	\mathcal{E}[m_{1}(\cdot|w_{0})](y_{0}) \} + \gamma  E_{Y_{1}|Y_{0}} \left[ m_{1}(Y_{1}|w_{0}) \left( E_{X}[\mathbb{U}_{1}(\mathbf{c}^{L};w_{0},X_{1},Y_{1})]  \right) \mid y_{0} \right],
	\end{align*}
	for any $m_{1}$ that satisfies the restrictions imposed in the text; it thus holds, in particular, for the value that attains the minimum. Note that this holds for any $(t,y^{t})$, not just $(t=0,y_{0})$, i.e.,
	\begin{align}
	\mathbb{U}_{t}(\mathbf{c}^{L};w^{t}) \leq & \min_{m_{t+1}} \left\{ c^{L}_{t}(w^{t}) + \theta \gamma
	\mathcal{E}[m_{t+1}(\cdot|w^{t})](y_{t}) \right. \notag\\
	&\left.+ \gamma E_{Y_{t+1}|Y_{t}}[ m_{t+1}(Y_{t+1}|w^{t}) \left( E_{X}[\mathbb{U}_{t+1}(\mathbf{c}^{L};w^{t},X_{t+1},Y_{t+1})]  \right) \mid y_{t} ]  \right\}. \label{eqn:BE_MA_2}
	\end{align}
	Therefore, putting together equations (\ref{eqn:BE_MA_1}) and (\ref{eqn:BE_MA_2}), it follows that $(\mathbb{U}_{t})_{t}$ satisfy the FP-MA.\\
	
	(b) Let $(\bar{U}_{t})_{t}$ satisfy the FP-MA and equation (\ref{eqn:TC_MA}). Then, by a simple iteration it is easy to see that
	\begin{align*}
	\bar{U}_0(\mathbf{c}^{L};w_{0}) \leq & \lim_{T \rightarrow \infty} \sum_{j=0}^{T}
	\gamma^{j} E_{W^{j}|W_{0}} \left[ \left( M_{j}(W^{j}) \right) \{  c^{L}_{j}(W^{j})  + \theta \gamma
	\mathcal{E}[m_{j+1}(\cdot|W^{j})](Y_{j}) \} \mid w_{0} \right]\\
	& + \lim_{T \rightarrow \infty} \gamma^{T+1} E_{W^{T+1}|W_{0}}[ M_{T+1}(W^{T+1}) \left( \bar{U}_{T+1}(\mathbf{c}^{L};W^{T+1})  \right) |w_{0}].
	\end{align*}
	The last term in the RHS is zero by equation (\ref{eqn:TC_MA}), so $\bar{U}_0(\mathbf{c}^{L};w_{0}) \leq \mathbb{U}_0(\mathbf{c}^{L};w_{0})$ (where $\mathbb{U}$ satisfies the SP-MA). The reversed inequality follows from similar arguments and the fact that $\mathbb{U}_0(\mathbf{c};w_{0})$ is the minimum possible value.
	
	The proof for $(t,y^{t})$ is analogous. Therefore, we conclude that any sequence of functions $(\bar{U}_{t})_{t}$ that satisfies FP-MA and (\ref{eqn:TC_MA}), also satisfies SP-MA.

	\bigskip
	
	\textbf{Proof of $\star$.}  To show $\star$ is valid, let
	\begin{align*}
	H_{n} \equiv \sum_{s=0}^{n} m_{1}(y_{1}|w_{0})
	\gamma^{s} E_{W^{s+1}|W^{1}} \left[ \frac{M_{s+1}(W^{s+1})}{M_{1}(w^{1})} \{  c^{L}_{s+1}(W^{s+1})  + \theta \gamma
	\mathcal{E}[m_{s+2}(\cdot|W^{s+1})](Y_{s+1}) \}  \mid w^{1} \right] .
	\end{align*}
	We note that
	\begin{align*}
	|H_{n}| \leq & \sum_{s=0}^{\infty} m_{1}(y_{1}|w_{0}) \gamma^{s}  C E_{W^{s+1}|W^{1}} \left[ \frac{M_{s+1}(W^{s+1})}{M_{1}(w^{1})} \mid w^{1} \right] \\
	& + \sum_{s=0}^{\infty} m_{1}(y_{1}|w_{0}) \gamma^{s} E_{W^{s+1}|W^{1}} \left[ \frac{M_{s+1}(W^{s+1})}{M_{1}(w^{1})}  \theta \gamma
	\mathcal{E}[m_{s+2}(\cdot|W^{s+1})](Y_{s+1}) \mid w^{1} \right] ,
	\end{align*}
	where the second line follows because $c_{t}^{L}$ is bounded. Observe that, $E_{W^{j+1}|W^{t}} \left[ \frac{M_{j+1}(W^{j+1})}{M_{t}(w^{t})} \mid w^{t} \right] = 1$ for all $t$ and $j$ and $\sum_{s=0}^{\infty} \gamma^{s} E_{W^{s+1}|W^{1}} \left[ \frac{M_{s+1}(W^{s+1})}{M_{1}(w^{1})}  \theta \gamma
	\mathcal{E}[m_{s+2}(\cdot|W^{s+1})](Y_{s+1}) \mid w^{1} \right] \leq C_{C,\gamma,\theta}$ by Lemma \ref{lem:min-M}. Hence
	\begin{align*}
	|H_{n}| \leq m_{1} \times K_{0}
	\end{align*}
	for some $ \infty >K_{0}>0$ (it depends on $(\gamma,\theta,M)$). Since the RHS is integrable, by the Dominated Convergence Theorem, interchanging summation and integration is valid.
\end{proof}

\begin{proof} \textsc{Proof of Lemma \ref{lem:min-M}.}
	Before showing the desired results, we show it suffices to perform the minimization over $(m_{t})_{t} \in \mathbb{M}$ where
	\begin{align*}
	\mathbb{M} \equiv \left\{(m_{t})_{t} \colon m_{t} \in \mathcal{M} \cap \sum_{j=0}^{\infty}
	\gamma^{j}   E \left[ \left( \frac{M_{t+j}(W^{t+j})}{M_{t}(w^{t})} \right)
	\mathcal{E}[m_{t+j+1}(\cdot|W^{t+j})](Y_{t+j})  \} | w^{t} \right] \leq C_{C,\gamma,\theta}, \forall y^{t} \right\},
	\end{align*}
	where $C_{C,\gamma,\theta} = 2\frac{C}{(1-\gamma)\theta \gamma }$.
	
	We do this by contradiction. Suppose that $(m_{t})$ solves the minimization problem in SP-MA, and, $ \sum_{j=0}^{\infty}
	\gamma^{j}   E \left[ \left( \frac{M_{t+j}(W^{t+j})}{M_{t}(w^{t})} \right)
	\mathcal{E}[m_{t+j+1}(\cdot|W^{t+j})](Y_{t+j})  \} | w^{t} \right] > C_{C,\gamma,\theta}$. Since consumption is bounded
	\begin{align*}
	\mathbb{U}_{t}(\mathbf{c}^{L};w^{t}) = & \sum_{j=0}^{\infty}
	\gamma^{j}   E \left[ \left( \frac{M_{t+j}(W^{t+j})}{M_{t}(w^{t})} \right)
	\{ c^{L}_{t+j}(W^{t+j}) + \gamma \theta \mathcal{E}[m_{t+j+1}(\cdot|W^{t+j})](Y_{t+j})  \} | w^{t} \right] \\
	\geq &  \sum_{j=0}^{\infty}
	\gamma^{j}   \left\{ (-C) E \left[  \left( \frac{M_{t+j}(W^{t+j})}{M_{t}(w^{t})} \right) \mid w^{t} \right] \right. \\
	&\left.+  \theta \gamma E \left[ \left( \frac{M_{t+j}(W^{t+j})}{M_{t}(w^{t})} \right) \mathcal{E}[m_{t+j+1}(\cdot|W^{t+j})](Y_{t+j}) \mid w^{t} \right]  \right\}.
	\end{align*}
	Note that
	{\footnotesize{
			\begin{align*}
			E \left[  \left( \frac{M_{t+j}(W^{t+j})}{M_{t}(w^{t})} \right) \Big| w^{t} \right] &= \int_{\mathbb{W}^{t+j-1}|\mathbb{W}^{t} } \hspace{-0.2in}\frac{M_{t+j-1}(\omega^{t+j-1})}{M_{t}(w^{t})}  \left\{ \int_{\mathbb{Y}} m_{t+j}(\omega^{t+j-1}\hspace{-0.08in},y_{t+1}) P(dy_{t+j}|y_{t+j-1}) \right\} Pr(d\omega^{t+j-1}|w^{t})\\
			&= \int_{\mathbb{W}^{t+j-1}|\mathbb{W}^{t} } \frac{M_{t+j-1}(\omega^{t+j-1})}{M_{t}(w^{t})} Pr(d\omega^{t+j-1}|w^{t}) = ... = 1.
			\end{align*}}}
	where $Pr$ is the conditional probability over histories $W^{\infty}$, given $W^{t} = w^{t}$.
	Hence,
	\begin{align*}
	& \sum_{j=0}^{\infty}
	\gamma^{j}   E \left[ \left( \frac{M_{t+j}(W^{t+j})}{M_{t}(w^{t})} \right)
	\{ c^{L}_{t+j}(W^{t+j}) + \gamma \theta \mathcal{E}[m_{t+j+1}(\cdot|W^{t+j})](Y_{t+j})  \} | w^{t} \right] \\
	\geq & - \frac{C}{1-\gamma} +   \theta \gamma E \left[ \left( \frac{M_{t+j}(W^{t+j})}{M_{t}(w^{t})} \right) \mathcal{E}[m_{t+j+1}(\cdot|W^{t+j})](Y_{t+j}) \mid w^{t} \right].
	\end{align*}
	By assumption, the second term is larger than $\theta \gamma C_{C,\gamma,\theta}$. Hence, the value for the minimizing agent of playing $(m_{t})_{t}$  is bounded below by $- \frac{C}{1-\gamma} +  \theta \gamma C_{C,\gamma,\theta}$. By our choice of $C_{C,\gamma,\theta}$,
	\begin{align*}
	- \frac{C}{1-\gamma} +  \theta \gamma C_{C,\gamma,\theta} = \frac{C}{1-\gamma}.
	\end{align*}
	Since $\mathcal{E}[1] = 0$, the RHS of the previous display is larger than
	\begin{align*}
	\sum_{j=0}^{\infty}
	\gamma^{j} E \left[ \left( \frac{M_{t+j}(W^{t+j})}{M_{t}(w^{t})} \right) \{  c^{L}_{t+j}(W^{t+j})  + \theta \gamma
	\mathcal{E}[1](Y_{t+j}) \} \mid w^{t} \right].
	\end{align*}
	Therefore, we conclude that
	\begin{align*}
	\mathbb{U}_{t}(\mathbf{c}^{L};w^{t}) >  \sum_{j=0}^{\infty}
	\gamma^{j} E \left[ \left( \frac{M_{t+j}(W^{t+j})}{M_{t}(w^{t})} \right) \{  c^{L}_{t+j}(W^{t+j})  + \theta \gamma
	\mathcal{E}[1](Y_{t+j}) \} \mid w^{t} \right].
	\end{align*}
	But since $m_{t}=1$ for all $t$ is a feasible choice, this is a contradiction to the definition of $\mathbb{U}_{t}(\mathbf{c}^{L};w^{t})$.
\end{proof}

\section{Moments of Approximating and Distorted Densities.}
\label{app:mom}

Table \ref{tab:MOM} presents the computed moments for the approximating and distorted conditional densities of next-period $y_{t+1}$, given current $y_{t}$ and bond holdings $B_{t}$. As shown in Figure \ref{fig:2_densities}, the current endowment level $y_{t}$ is set to half a standard deviation below its unconditional mean, and the bond holdings $B_{t}$ is given by the median of its unconditional distribution in the simulations.
\begin{table}[htp] 
  \centering
  \begin{tabular}{l|c|p{0.05cm}c}
   \hline\hline
Moment & Approximating Model &  & Distorted Model \\ \hline
Mean($y_{t+1}$) & $0.9518$  &  & $0.9481$ \\
Std.dev.($y_{t+1}$) & $0.0191$  &  & $0.0202$ \\
Skewness($y_{t+1}$) & $0.0601$  & & $0.0811$ \\
Kurtosis($y_{t+1}$) & $3.0064$ & & $2.7910$ \\ \hline\hline
  \end{tabular}
\caption{Moments for the approximating and distorted conditional densities.}
\label{tab:MOM}
\end{table}

By Law of Large Numbers, the moments for the approximating model are essentially the same to the corresponding ``population'' moments of the lognormal distribution. Regarding the distorted model, several moments differ significantly from those of the approximating model. In particular, there is a clear shift to the left of the conditional mean. Because of it, the skewness is higher, even though the distorted model puts more probability mass on low realizations of output, $y_{t+1}$, where default is optimal for the borrower, as illustrated in Figure \ref{fig:2_densities}. 

\section{Micro-foundations for ad-hoc Pricing Kernels}
\label{sec:kernel}

In recent years, several studies on quantitative sovereign default models have considered ad-hoc pricing kernels to improve the calibration along the asset-pricing dimension while keeping the model tractable and easy to solve. Some examples include \cite{ARELLANO_AER08}, \cite{Arellano_WP}, and \cite{Hatchondo_Debt_Dilution}. We view our model as providing foundations for this class of ad-hoc pricing kernels. In this section, we study the differences and similarities between the our and the ad-hoc pricing kernels, both theoretically and quantitatively.

In the aforementioned papers, the pricing kernel is given by an ad-hoc function that belongs to the class $\mathcal{S}$ defined by
\begin{equation}
	\mathcal{S}\equiv \left\{S:\mathbb{Y\times Y} \rightarrow \mathbb{R_{++}} \text{ such that } E [S(y_{t},Y_{t+1})|y_{t}] = \gamma \text{ and } S(y_{t},\cdot) \text{ is non-increasing} \right\},
\end{equation}
where $\gamma$ is the lenders' time discount factor, which is equal to the reciprocal of the gross risk-free rate, i.e. $\gamma=\frac{1}{1+r^{f}}$. Note that $S(y_{t},\cdot)$ scaled up by $\frac{1}{\gamma}$, i.e. $\frac{S(y_{t},\cdot)}{\gamma}$, is a pdf on $\mathbb{Y}$. In what follows we assume that $Y$ has a pdf, denoted by $f_{Y'|Y}$, and that the pdf embedded in $S(y_{t},\cdot)$ and $f_{Y'|Y}$ are equivalent.\footnote{Two probability measures are equivalent if they are absolutely continuous with respect to each other.}

A common example is
\begin{equation}
	S(y_{t},y_{t+1}) = \gamma \exp \{- \eta \upsilon_{t+1}-0.5(\eta \sigma_{\upsilon})^2\}, \label{asdf}
\end{equation}
where $\eta > 0$, $\upsilon_{t+1} \sim N(0,\sigma^{2}_{\upsilon})$, and the endowment of the borrower follows an AR(1),
\begin{align}\label{eqn:SDF-AR}
	\log y_{t+1} = \alpha + \rho \log y_{t} + \upsilon_{t+1}.
\end{align}
This process is typically assume in the literature (it is used in our simulation results as well) and facilitates the exposition.

It is easy to see that the equilibrium price function associated to an ad-hoc pricing kernel $S \in \mathcal{S}$ and an arbitrary stochastic process for output with pdf $f_{Y'|Y}$ is given by
\begin{align*}
	B' \mapsto q_{a}(y,B') = E_{Y} [ \{ \lambda + (1-\lambda) (\psi+ q_{a}(y',B^{\ast}_{a}(y',B'))) \} \delta^{\ast}_{a}(y',B') S(y,y^{\prime})|y],
\end{align*}
for any $(y,B') \in \mathbb{Y}\times \mathbb{B}$; where $E_{Y}[\cdot|y]$ is computed under pdf $f_{Y'|Y}$, and $B^{\ast}_{a}$ and $\delta^{\ast}_{a}$ denote the equilibrium debt and default policies, respectively, given pricing kernel $S$.

Due to the equivalence assumption, it is easy to see that
\begin{align} \label{eqn:price-SDF}
	q_{a}(y,B') = \gamma \int_{\mathbb{Y}} \{ \lambda + (1-\lambda) (\psi+ q_{a}(y',B^{\ast}_{a}(y',B')) \} \delta^{\ast}_{a}(y',B')\varphi(y'|y) dy',
\end{align}
where $\varphi(\cdot|y)$ is a new pdf that depends on the primitives of pdf $f_{Y'|Y}$ and parameters of $S$. That is, using this ad-hoc pricing kernel is equivalent to using a modified version of the conditional probability governing the stochastic process of the endowment. Moreover, it can be shown that for any $S \in \mathcal{S}$, $\varphi(\cdot|y)$ is first order dominated by $f_{Y'|Y}(\cdot|y)$.
\footnote{We view this as a noteworthy similarity with our model pricing kernel, $\gamma m^{\ast}_{R}$, which also results on a pricing equation that uses a distorted version of the conditional probability governing the stochastic process of the endowment. Our model pricing kernel, however, emerges endogenously in general equilibrium from the lenders' attitude towards model uncertainty, and this fact has important consequences. First, our conditional distorted probability is not Markov, i.e., depends on the entire past history (as opposed to only depending on last period value) of endowment. This is due to the fact that our conditional distorted probability depends on $B_{t+1}^{\ast}$ (and access to financial markets), whereas the probability measure in equation (\ref{eqn:price-SDF}) does not.}

We finally observe that for our previous example with the kernel specification (\ref{asdf}) and output process (\ref{eqn:SDF-AR}), $\varphi(\cdot|y)$ is a log-normal pdf with parameters $(-\sigma^{2}_{\upsilon} \eta + \alpha + \rho \log y,\sigma_{\upsilon}^{2})$.
In particular, the conditional probability used in the pricing equation is still \emph{log-normal with the same variance but with lower conditional mean}; that is, it is first order dominated by the one governing the stochastic process of the endowment, and the parameter $\eta$ regulates how different these two distributions are. Observe, however, that even with the  output process (\ref{eqn:SDF-AR}), the conditional distorted probability measure in our model is not longer log-normal; in particular, it is skewed to the left, as shown in Figure \ref{fig:2_densities}.\footnote{Surprisingly, this modified pdf resembles the distorted pdf that emerges endogenously under model uncertainty in \cite{BHS} to analyze the equity premium and the risk-free rate puzzle in the context of \cite{Hansen_Jagannathan} bounds. Both for a random walk process and a trend stationary process for log consumption, the distorted pdf results as well from a conditional mean shift in the approximating one.}

	
	In order to shed further light on asset-pricing implications of the ad-hoc pricing kernel and our pricing kernel, we find convenient to work with the modified pdf and the distorted pdf and to assume that the default set is of the threshold type and the same across different pricing kernels. We also focus the analysis on the short-term debt model, i.e., $\lambda=1$. The assumption over the default sets being of the the same and of the threshold type, although is ad-hoc, it seems to hold true in the numerical simulations and also have been shown to hold in different environments for these type of models; see \cite{ARELLANO_AER08} and \cite{Pouzo}.	
	Formally, let $i\in \{ \eta,\theta\}$ where $\eta$ ($\theta$) denotes the economy with ad-hoc (ours) pricing kernel. Suppose the stochastic process for the endowment is given by equation (\ref{eqn:SDF-AR}), $\lambda = 1$ and $B' \mapsto D^{\ast}_{i}(B') = D^{\ast}(B') \equiv \{ y' : y' \leq \bar{y}(B')\}$, then, for all $B'$,  the spread can be constructed as follows:
	\begin{align*}
		\mathbf{Sp}_{i,t+1}(B') = \left( \gamma \int_{ y' > \bar{y}(B')}  f^{i}_{t+1}( y' | y^{t+1})  \right)^{-1} - \gamma^{-1} 	= \gamma^{-1} \frac{F^{i}_{t+1}(\bar{y}(B')| y^{t+1})}{1- F^{i}_{t+1}(\bar{y}(B')| y^{t+1})},
	\end{align*}
	where $f^{i}_{t}(\cdot | y^{t})$ ($F^{i}_{t}(\cdot | y^{t})$) is the conditional pdf (cdf) of the model $i$ given history $y^{t}$.
	
	It follows that, if for a given $y_{t}$,
	\begin{align}\label{eqn:wedge}
		F^{\eta}_{t}(\cdot | y^{t}) > (<) F^{\theta}_{t}(\cdot | y^{t})
	\end{align}
	holds, then $\mathbf{Sp}_{\eta,t+1}(B') > (<) \mathbf{Sp}_{\theta,t+1}(B')$ a.s..
	
	
	A few remarks regarding this result are in order. First, observe that for states with high values of endowment, as shown in the bottom panels of Figure \ref{fig:3_densities_yb}, our conditional distorted pdf is well-approximated by $F_{Y'|Y}$--- i.e., distortions are negligible ---, consequently, we expect equation (\ref{eqn:wedge}) to hold with the ``$<$'' inequality; i.e., our conditional cdf $F_{t}^{\theta}$ dominates (in first order stochastic sense) the cdf corresponding to $F_{t}^{\eta}$. We then conclude that for these states, our model generates an spread that is lower than then one generated by the model with ad-hoc pricing kernel. On the other hand, for states with low endowments, we expect our distorted conditional probability measure to put more weight on low values of future endowment than $F_{t}^{\eta}$; e.g. see the top panels of Figure \ref{fig:3_densities_yb}, so the inequality in equation should be reversed and therefore our model generates an spread that is higher than then one generated by the model with ad-hoc pricing kernel.

\section{Robustness Checks}
\label{sec:robcheck}\label{app:RA}

In this section we present robustness checks. \\

\textbf{Different degrees of concern about model misspecification.} In Table \ref{tab:rob} we report some business cycle statistics from the simulations of
our model for different degrees of model uncertainty and no risk aversion on
the lenders' side. We start with no fears about model
misspecification, i.e. $\theta =+\infty ,$ and we lower the penalty parameter
to $0.25$, for which we obtain a detection error probability of $9.5$
percent. As expected when we reduce the value of $\theta$, we observe that the frequency
of default goes down. To keep it at the historical level of $3$ percent per year, we make the borrower more impatient by adjusting $\beta$ downwards.

In the comparison across models, which typically differ in several dimensions along their parametrization and assumed functional forms, it may be hard to identify which key ingredient is driving each difference in the simulated statistics. Table \ref{tab:rob} helps us highlight the contribution of model uncertainty by showing how the dynamics of relevant macro variables vary in the same environment as we increase the preference for robustness.

\begin{table}[ht]
	\centering
	\begin{tabular}{l|c|ccccc}
		\hline\hline
		Statistic & $\theta =+\infty $ & $\theta =5$ & $\theta =1$ & $\theta =0.75$ &
		$\theta =0.5$ & $\theta =0.25$ \\ \hline
		Mean$(r-r^{f})$ & $4.54$ & $4.83$ & $6.43$ & $7.23$ & $9.30$ & $18.74$ \\
		Std.dev.$(r-r^{f})$ & $3.32$ & $3.41$ & $3.96$ & $4.28$ & $5.10$ & $9.36$\\
		Mean$(-b/y)$ & $43.32$ & $43.40$ & $43.93$ & $43.96$ & $43.78$ & $42.22$\\
		Std.dev.$(c)/$std.dev.$(y)$ & $1.17$ & $1.18$ & $1.21$ & $1.22$ & $1.23$ & $1.22$ \\
		Std.dev.$(tb/y)$ & $0.86$ & $0.92$ & $1.11$ & $1.17$ & $1.28$ & $1.40$\\
		Corr$(y,c)$ & $0.99$ & $0.99$ & $0.98$ & $0.98$ & $0.98$ & $0.97$ \\
		Corr$(y,r-r^{f})$ & $-0.79$ & $-0.78$ & $-0.77$ & $-0.76$ & $-0.74$ & $-0.67$ \\
		Corr$(y,tb/y)$ & $-0.77$ & $-0.76$ & $-0.72$ &$-0.70$ & $-0.66$ & $-0.55$ \\
		& \multicolumn{1}{|l|}{} &  &  &  &  &  \\
		DEP & $0.50$ & $0.469$ & $0.377$ & $0.344$ & $0.267$ & $0.095$ \\
		& \multicolumn{1}{|l|}{} &  &  &  &  &  \\
		Default frequency (annually) & $3.00$ & $3.00$ & $3.00$ & $3.00$ & $3.00$ & $3.00$ \\
		\hline\hline
	\end{tabular}
	\caption{Business Cycle Statistics for Different
		Degrees of Robustness}
	\label{tab:rob}
\end{table}

The first feature that stands out is that both the mean and the standard deviation of bond spreads increase with the lenders' concerns about model misspecification. For the same default frequency, as $\theta $ decreases, a greater degree of concern about model misspecification tends to push up the probability distortions associated with low utility states for the lender, in particular those states in which default occurs.

Note, however, that on average borrowing almost does not decline as its cost goes up. While this is true for long-term debt, it is not when the borrower can only issue one-period bonds. In the latter case, the borrower adjusts much more its debt level as output slides down and default risk (under both models) increases. This follows from the fact that, the disincentives to issue an additional bond are larger with one-period bonds that with long-term debt.\footnote{In the working paper version, \cite{Pouzo_Presno} report the MC statistics for different degrees of concern about model misspecification for one-period debt.}

Given that borrowing does not adjust enough to compensate for prices variations, interest repayments become more volatile as $\theta$ decreases. Consequently, we observe more variability in trade balance and consumption relative to output.

	\bigskip

\textbf{Risk aversion with time-additive, standard expected utility.} As displayed in Table \ref{tab:RA}, plausible degrees of risk aversion on the lenders' side with standard time separable expected utility, are not enough to generate sufficiently high bond spreads while keeping the default
frequency as observed in the data.

\begin{table}[ht]
  \centering
  \begin{tabular}{l|p{0.2cm}p{1.4cm}p{1.4cm}p{1.4cm}p{1.4cm}p{1.4cm}p{1.4cm}}
     \hline\hline
Statistic &  & \small{$\sigma ^{L}=1$} & \small{$\sigma ^{L}=2$} & \small{$\sigma^{L}=5$} & \small{$\sigma^{L}=10$} & \small{$\sigma^{L}=20$} & \small{$\sigma^{L}=50$}\\ \hline
Mean$(r-r^{f})$ &  & $3.44$ & $3.48$ & $3.48$                 & $3.49$ & $3.49$ & $3.76$ \\
Std.dev.$(r-r^{f})$ &  & $2.61$ & $2.63$ & $2.59$             & $2.61$ & $2.61$ & $2.82$ \\
Mean$(r^{f})$ &  & $4.05$ & $4.04$ & $3.98$                   & $3.79$ & $3.05$ & $-1.38$ \\
Std.dev.$(r^{f})$ &  & $0.19$ & $0.39$ & $0.96$               & $1.91$ & $3.83$ & $9.12$ \\
Mean$(-b/y)$ &  & $75.49$ & $74.93$                           & $74.51$ & $74.84$ & $75.36$  & $77.46$ \\
Std.dev.$(c)/$std.dev.$(y)$ &  & $2.32$ & $2.31$ & $2.31$     & $2.33$ & $2.36$ & $2.66$ \\
Std.dev.$(tb/y)$ &  & $6.65$ & $6.64$ & $6.63$                & $6.69$ & $6.85$ & $7.96$ \\
Corr$(y,c)$ &  & $0.70$ & $0.70$ & $0.70$                     & $0.70$ & $0.69$ & $0.64$ \\
Corr$(y,r-r^{f})$ &  & $-0.60$ & $-0.61$ & $-0.61$            & $-0.60$ & $-0.61$ & $-0.57$\\
Corr$(y,tb/y)$ &  & $-0.36$ & $-0.36$ & $-0.36$               & $-0.36$ & $-0.36$ & $-0.33$ \\
&  &  &  &  \\
Default frequency (annually) &  & $3.00$ & $3.00$ & $3.00$               & $3.00$ & $3.00$ & $3.00$ \\ \hline\hline
  \end{tabular}
  \caption{Business Cycle Statistics for Different
Degrees of Risk Aversion.}
\label{tab:RA}
\end{table}

We considered an exogenous stochastic process for the lenders'
consumption given by%
\begin{equation*}
\ln C_{t+1}^{L}=\rho^{L} \ln C_{t}^{L}+\sigma_{\varepsilon} ^{L}\varepsilon _{t+1}^{L},
\end{equation*}%
where $\varepsilon _{t+1}^{L}\sim i.i.d.\mathcal{N}(0,1).$ Shocks $%
\varepsilon _{t+1}$ and $\varepsilon _{t+1}^{L}$ are assumed to be
independent. We estimate the log-normal AR(1) process for $C^{L}_{t}$ using U.S. consumption data.\footnote{Time series for seasonally adjusted real consumption of nondurables and services at a quarterly frequency are taken from the Bureau of Economic Analysis, in logs, and filtered with a linear trend. The estimates for parameters $\rho^{L}$ and $\sigma_{\varepsilon} ^{L}$ are $0.967$ and $0.025$, respectively.}

Table \ref{tab:RA} displays the business cycle statistics for different values of the lenders' coefficient of relative risk aversion, $\sigma^{L}$, ranging from 1 to 50, and no fears about model misspecification, i.e. $\theta =+\infty$.\footnote{%
For each value of $\sigma ^{L},$ the discount factor for the borrower, $%
\beta,$ is calibrated to replicate a default frequency of $3$ percent
annually.} First, as we would expect, bond spreads increase on average and become more volatile with the value of $\sigma^{L}$. They do so, however, to a very limited extent. Plausible degrees of risk aversion are not even close to generate sufficiently high bond spreads while keeping the default
frequency as observed in the data. Setting $\sigma^{L}$ equal to 50 generates average bond spreads of just $3.76$ percent, less than half the value observed in the data. This high value for the coefficient of risk aversion is sufficient to explain the equity premium puzzle in \cite{Mehra_Prescott}. In contrast with the economy considered there, the stochastic discount factor in our model would not typically vary inversely with the bond payoff, limiting the ability of the model to generate sufficiently high bond spreads.
Second, given a stationary process for consumption in our model, the net risk-free rate decreases and turns negative for sufficiently high values of $\sigma^{L}$, while its volatility grows dramatically. Facing lower risk-free rates, the borrower reacts by borrowing more. The variations in the debt-to-output ratio are, however, small, as in the case with model uncertainty.
Finally, larger and more volatile capital outflows for interest payments translate into higher variability of consumption and net exports.

\end{document}